\documentclass[10pt,a4paper]{article}

\usepackage[
    top=20mm,
    bottom=20mm,
    left=18mm,
    right=18mm
]{geometry}

\usepackage{graphicx}
\usepackage{multicol}
\usepackage{titlesec}
\usepackage{fancyhdr}
\usepackage{setspace}
\usepackage{array}
\usepackage{booktabs}
\usepackage{tikz}
\usepackage{changepage}
\usepackage{caption}
\usepackage{fontspec}
\usepackage{amsmath}
\usepackage{lipsum}
\usepackage{varwidth}
\usepackage{adjustbox}

\usepackage[english]{babel}

\usepackage{url}
\usepackage{amsfonts}
\usepackage{nicefrac}
\usepackage{microtype}
\usepackage{multirow}
\usepackage{tabularx}
\usepackage{lineno}

\usepackage[table,xcdraw,dvipsnames]{xcolor}

\newcolumntype{Y}{>{\raggedright\arraybackslash}X}

\usepackage{siunitx}

\usepackage[most]{tcolorbox}
\tcbuselibrary{breakable}

\definecolor{HeaderGray}{gray}{0.90}
\definecolor{OurModelLight}{gray}{0.97}
\definecolor{OurModelDark}{gray}{0.93}
\definecolor{lavender}{RGB}{216,210,245}
\definecolor{lightlavender}{RGB}{240,242,255}
\definecolor{bestblue}{RGB}{54,85,228}

\definecolor{NRboxbg}{RGB}{245,245,245}
\definecolor{NRboxframe}{RGB}{200,200,200}

\definecolor{headerblue}{RGB}{0,51,102}
\definecolor{rowblue}{RGB}{240,245,255}
\definecolor{tether-gray}{HTML}{F2F1EF}

\usepackage{hyperref}

\hypersetup{
    colorlinks=true,
    linkcolor=bestblue,
    citecolor=bestblue,
    filecolor=magenta,
    urlcolor=cyan
}

\newfontfamily\sharpfont[
    Path=./,
    BoldFont=SharpGrotesk-Medium20.otf,
    ItalicFont=SharpGrotesk-BookItalic20.otf,
    BoldItalicFont=SharpGrotesk-BoldItalic20.otf
]{SharpGrotesk-Book20.otf}

\DeclareCaptionFont{sharp}{\sharpfont}

\titleformat{\section}
    {\sharpfont\large\bfseries}
    {\thesection}
    {1em}
    {}

\titleformat{\subsection}
    {\sharpfont\normalsize\bfseries}
    {\thesubsection}
    {1em}
    {}

\titleformat{\subsubsection}
    {\sharpfont\normalsize\bfseries}
    {\thesubsubsection}
    {1em}
    {}

\titleformat{\paragraph}[runin]
    {\sharpfont\normalsize\bfseries}
    {\theparagraph}
    {1em}
    {}

\newcommand{\captiontitle}[1]{{\sharpfont\bfseries #1}}

\newtcolorbox{reviewbox}[2][]{%
    enhanced,
    breakable,
    colback=tether-gray,
    colframe=tether-gray,
    boxrule=0pt,
    left=15pt,
    right=15pt,
    top=15pt,
    bottom=15pt,
    fonttitle=\sharpfont\bfseries,
    coltitle=black,
    title=#2,
    #1
}

\renewtcolorbox[
  auto counter
]{reviewbox}[2][]{%
  title={Box~\thetcbcounter\ | #2},
  label={#1},
  breakable,
  colback=tether-gray,
  colframe=tether-gray,
  boxrule=0pt,
  arc=5pt,
  outer arc=5pt,
  left=15pt,
  right=15pt,
  top=15pt,
  bottom=15pt,
  toptitle=6pt,
  bottomtitle=4pt,
  coltitle=black,
  fonttitle=\sharpfont\bfseries
}

\begin{document}


\begin{flushleft}
\includegraphics[height=20pt]{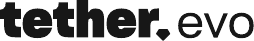}
\end{flushleft}

\vspace{10pt}


\begin{tcolorbox}[
    colback=tether-gray,
    colframe=tether-gray,
    boxrule=0pt,
    arc=5pt,
    left=15pt,
    right=15pt,
    top=15pt,
    bottom=15pt,
    width=\textwidth
]


{\sharpfont\huge\bfseries
From Neurons to Conversation: Speech Brain-Computer Interfaces
}

\vspace{5pt}


{\small\bfseries
Moein Khajehnejad\textsuperscript{1,2,}\footnote{The authors contributed equally to this manuscript.},
Forough Habibollahi\textsuperscript{1,a},
Tommaso Boccato\textsuperscript{1},
Margarida Sousa\textsuperscript{1},
Michal Olak\textsuperscript{1},
Francesco Jamal Sheiban\textsuperscript{1},
Matteo Ferrante\textsuperscript{1,3}

}

\vspace{2.5pt}

{\small
\textsuperscript{1}Tether Evo, 
\textsuperscript{2}Turner Institute for Brain and Mental Health, Monash University
\textsuperscript{3}The University of Rome Tor Vergata,
}

\vspace{15pt}


{\footnotesize
Speech brain--computer interfaces (BCIs) aim to restore communication by transforming neural activity related to speech, language, or communicative intent into external outputs such as text, synthesized voice, or avatar control. Recent advances in intracortical and electrocorticographic recording, deep sequence models, and language-model-assisted decoding have enabled rapid progress, including high-performance attempted-speech decoding and increasingly naturalistic speech synthesis. Yet these achievements also reveal that speech BCIs are not simply neural-to-text decoders. They are adaptive clinical systems in which neural representations, recording hardware, decoding architectures, language priors, feedback, and user learning interact over time.

In this Review, we synthesize speech BCI research from a system-level perspective. We first examine the neural substrates of speech and language, emphasizing their hierarchical, distributed, temporally structured, and non-stationary organization. We then examine recording and decoding choices, closed-loop adaptation, evaluation, clinical translation, and ethics. Across these domains, we highlight recurring trade-offs between signal resolution and invasiveness, low-level motor and high-level semantic targets, decoder accuracy and user agency, and language-model fluency and faithful neural evidence.

We argue that the next generation of speech BCIs should be evaluated not only by offline accuracy, but also by robustness across sessions, calibration burden, latency, uncertainty, usability, and safeguards against unintended decoding. By reframing speech BCIs as adaptive, user-centred systems, we outline the interdisciplinary priorities spanning speech neuroscience, neural engineering, machine learning, clinical practice, and neuroethics needed to move from proof-of-concept decoding toward reliable, expressive, and controllable communication neuroprostheses.

}

\end{tcolorbox}

\vspace{10pt}




\section{Introduction}

\subsection{Motivation and clinical need}

Speech enables rapid, flexible, and context-sensitive communication. Neurological conditions that disrupt speech production or articulation, including amyotrophic lateral sclerosis, brainstem stroke, spinal cord injury, and advanced neurodegenerative disease, can therefore cause profound loss of autonomy and quality of life despite largely preserved cognition \cite{Birbaumer2006BCIClinical,card2024accurate}. Restoring communication is consequently a central goal of brain--computer interface (BCI) research.

A speech BCI records neural activity, extracts task-relevant information, and translates it into text, synthesized voice, avatar control, or another communicative output, thereby bypassing impaired peripheral motor pathways. Its objective is not merely neural classification, but restoration of a communication channel that is sufficiently rapid, reliable, and controllable for functional use. Earlier BCIs established communication through cursor control or letter-by-letter selection, but generally remained limited in rate and usability \cite{Wolpaw2002BCIOverview,Farwell1988P300}. By accessing neural processes underlying speech planning and production, speech BCIs could support more fluent and naturalistic communication, particularly when high-resolution invasive recordings are clinically justified \cite{herff2015brain,brumberg2010brain}. Recent word-, sentence-, and continuous-text decoding studies demonstrate this potential \cite{moses2021neuroprosthesis,Willett2023SpeechBCI}, while also exposing persistent barriers: performance can degrade across time, require substantial training or recalibration, and generalize poorly across sessions and individuals \cite{Sussillo2016NeuralDrift,perge2014reliability}. Speech BCIs therefore serve both as prospective clinical tools and as experimental systems for studying speech, language, intention, and feedback-based learning.

\subsection{What constitutes "speech" for BCI?}

Speech is not represented by a single neural variable, but by distributed and temporally organized processes spanning articulatory motor plans, acoustic features, phonological units, lexical representations, semantics, prosody, and pragmatics \cite{Hickok2012SpeechNeuralBasis,Price2012LanguageBrain}. Speech BCI studies target different levels of this hierarchy: some reconstruct acoustic features or waveforms \cite{Pasley2012Reconstruction,Akbari2019SpeechReconstruction}; others decode phonemes, syllables, or words \cite{herff2015brain,Moses2019SentenceDecoding}; and others map neural activity directly to text or semantic representations \cite{zhang2025decoding,tang2023semantic}. These choices are not merely technical: they embody different assumptions about how speech is represented and which representations are accessible, stable, and useful for communication. They also determine the temporal resolution, vocabulary constraints, training labels, output flexibility, and error modes of the resulting system, with direct consequences for model design, interpretation, and generalization.

The term also encompasses overt, attempted, silently mouthed, imagined, and internally planned speech \cite{brumberg2010brain,martin2018decoding}. These paradigms differ in peripheral movement, motor intention, auditory--phonological imagery, and higher-level linguistic planning. Explicitly defining the behaviour and representational level being decoded is therefore essential for comparing studies and designing clinically meaningful systems.

\subsection{Why now: data scale, deep learning, chronic interfaces}

Recent progress reflects three converging developments. First, invasive recording technologies now provide higher-resolution cortical signals over longer periods, including chronic recordings and richer longitudinal datasets in clinical populations \cite{Collinger2013ChronicImplants}. Second, convolutional, recurrent, and attention-based sequence models have improved the extraction of high-dimensional spatiotemporal structure relative to earlier linear approaches \cite{Sussillo2016RNN,glaser2020machine}; self-supervised learning and speech- and text-pretrained foundation models further enable representation learning and transfer across tasks and users \cite{defossez2023decoding,jayalath2024brain}. Third, closed-loop experiments show that users adapt their neural strategies in response to decoder behaviour, causing neural representations to evolve in ways that can improve decoder performance over time \cite{Taylor2002LearningBCI,Ganguly2009Reorganization}.

\begin{figure}
    \centering
    \includegraphics[width=0.9\linewidth]{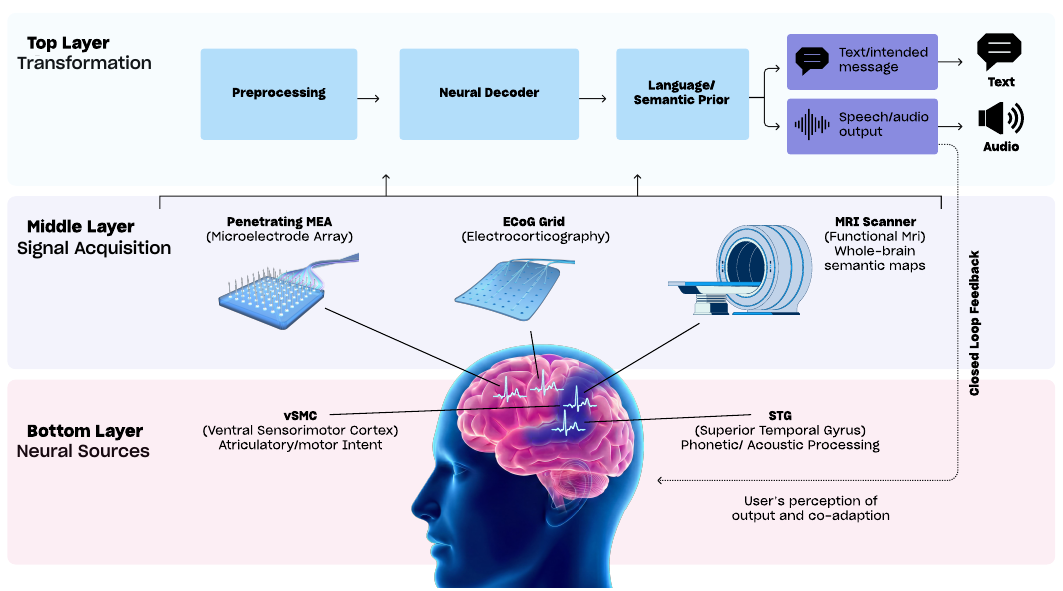}

    \caption{\captiontitle{Speech BCI system stack: from neural sources to closed-loop communication.}
    Speech BCIs can be viewed as a layered system linking neural sources, recording interfaces, decoding models, and user feedback.
    \textbf{Bottom layer:} neural sources and speech-related representations. Speech-related activity is distributed across cortical regions including ventral sensorimotor cortex (vSMC), associated with articulatory and motor-intent representations, and superior temporal gyrus (STG), associated with phonetic and acoustic processing.
    \textbf{Middle layer:} signal acquisition. Neural activity can be captured using recording technologies with different spatial, temporal, and clinical trade-offs, including penetrating microelectrode arrays (MEAs), electrocorticography (ECoG) grids, and non-invasive neuroimaging such as fMRI for whole-brain semantic mapping.
    \textbf{Top layer:} computational transformation. Recorded signals are preprocessed and passed through neural decoders, often combined with language or semantic priors, to generate communication outputs such as intended text or synthesized speech/audio.
    The dashed feedback pathway illustrates closed-loop operation, in which the user perceives the decoded output and adapts subsequent neural activity, strategy, and control over time.}
\label{fig:1}
\end{figure}

These developments motivate a systems-level view in which neural representations, recording hardware, decoding models, language priors, outputs, and feedback jointly determine performance. Figure~\ref{fig:1} summarizes this stack and the feedback processes through which communication becomes adaptive rather than purely feedforward.

\subsection{Scope and organization of this Review}

This Review synthesizes speech BCI research through a system-level framework integrating speech and language neuroscience, neural engineering, and contemporary machine learning. Rather than organizing the literature only by recording modality or decoding target, we focus on the common scientific and technical challenges that cut across speech BCI systems, including the hierarchical organization of speech, dynamic neural representations, mismatches between neural activity and decoding targets, and mutual adaptation between users and decoders.

We first examine the neural substrates and recording modalities that determine what information is available to a BCI, followed by representational targets and decoding paradigms ranging from linear methods to foundation models. We then consider closed-loop co-adaptation, evaluation, generalization, robustness, clinical translation, and ethical implications. Across these topics, we examine recurring trade-offs between signal resolution and invasiveness, low-level motor and higher-level linguistic targets, language-model fluency and fidelity to neural evidence, and decoder accuracy and user agency. We treat learning, stability, and failure as properties of the complete communication system rather than of the decoder alone. Box~\ref{box:comparison} summarizes how this framing extends prior reviews.


\begin{reviewbox}[box:comparison]{How this Review differs from prior speech BCI reviews}

Recent reviews of speech brain--computer interfaces (BCIs) have provided valuable summaries of experimental achievements across recording modalities, decoding targets, and model architectures \cite{brumberg2010brain,luo2022brain,silva2024speech}. However, much of the existing literature is organized around individual tasks (for example, phoneme- or word-level decoding) and emphasizes decoder performance under specific experimental conditions. This Review complements and extends these efforts by adopting a system-level perspective that highlights shared principles, limitations, and opportunities across the field.

\medskip
\noindent\textbf{Key distinctions of this Review include:}

\begin{itemize}
    \item \textbf{From study-by-study summaries to an integrated landscape.}  
    This Review provides an integrated synthesis of the current speech BCI landscape, bringing together decoding targets, datasets, and model architectures within a common framework and moves beyond a study-by-study description. We organize the field around shared computational and neural constraints imposed by the hierarchical, distributed, and temporally structured nature of speech representations in the brain.

  \item \textbf{From one-way decoding to closed-loop co-adaptation.}
    This Review moves beyond a predominantly feedforward view of speech BCIs and frames them as closed-loop neural--machine systems. In these systems, neural activity and decoding models are continuously shaped by feedback and user behaviour. We treat this mutual co-adaptation as a central organizing principle because it influences learning, stability across sessions, user experience, and long-term clinical viability.

    \item \textbf{Integration of modern machine learning paradigms.}  
    Beyond supervised deep learning, we examine the emerging role of self-supervised learning, foundation models, and multimodal representation alignment, not only as performance-enhancing tools but as conceptual bridges between neural computation and artificial language models.

    \item \textbf{Emphasis on generalization, robustness, and translation.}  
    We analyze why many speech BCI systems fail to generalize across sessions, users, or contexts, and discuss how representational mismatch and co-adaptive dynamics contribute to these limitations.

    \item \textbf{Expanded ethical and societal perspective.}  
    In addition to technical challenges, this Review addresses ethical considerations specific to speech BCIs, including mental privacy, user agency, authorship, provenance of decoded output, and emerging discussions around inner speech and neuro-rights.
\end{itemize}

\medskip
By reframing speech BCIs as adaptive, interactive systems rather than static decoders, 
this Review aims to provide a unifying framework that clarifies current bottlenecks 
and guides the development of robust, naturalistic speech neuroprostheses.

\end{reviewbox}

\section{Neural substrates of speech and language relevant to BCI}\label{neural_substrates}

Speech emerges from coordinated cortical and subcortical systems operating across multiple spatial and temporal scales rather than from a single locus or static code. A useful BCI account must therefore connect anatomy, representational level, population dynamics, and feedback instead of assigning speech to an isolated region. This organization constrains which variables can be decoded, how targets and labels should be defined, how neural signals should be interpreted, and why performance may fail to generalize across tasks, sessions, and individuals.

\subsection{Hierarchical organization of speech processing}

Speech processing is hierarchical, although the direction of information flow depends on behaviour. Perception generally progresses from acoustic analysis toward phonological, lexical, and semantic representations, whereas production transforms communicative intent into lexical, phonological, articulatory, and motor commands \cite{Goldstein2022Shared,Goldsteinetal2025,evanson2025emergencelanguagedevelopingbrain,gadonneix2026temporalstructurelanguagehierarchy,zhang2025thought}. Inner speech engages parts of this hierarchy without full muscular execution or sensory consequences. Accordingly, no single region represents ``speech'' in its entirety; different areas preferentially encode different levels of the speech-processing hierarchy \cite{Hickok2012SpeechNeuralBasis,Price2012LanguageBrain}.

\begin{figure}[!h]
    \centering
    \includegraphics[width=0.9\linewidth]{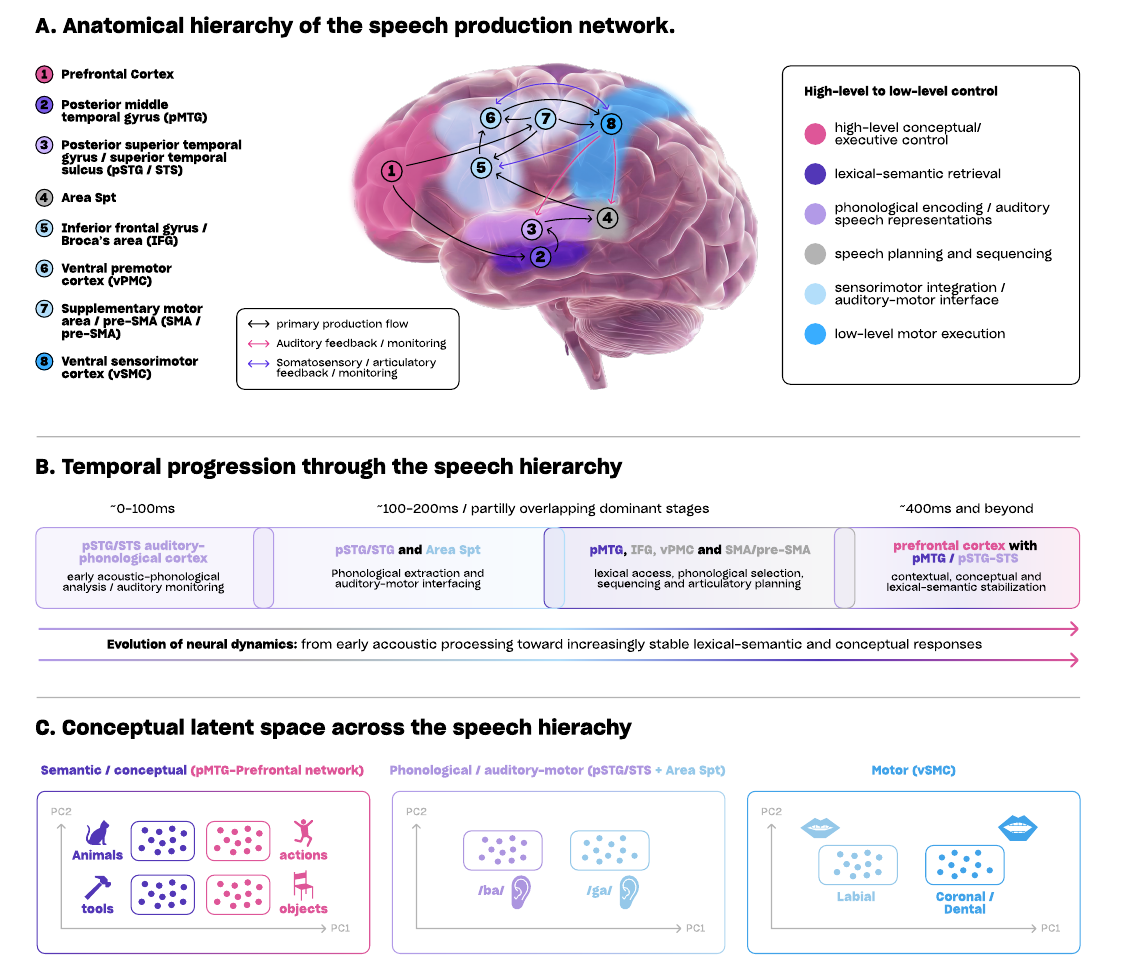}
    \caption{\captiontitle{Neuroanatomical, temporal, and representational structure of speech processing relevant to BCI.}
    \textbf{A.} Schematic anatomical hierarchy of the speech production network. Speech-related processing spans high-level conceptual and executive control, lexical--semantic retrieval, phonological and auditory speech representations, speech planning and sequencing, sensorimotor integration, and low-level articulatory motor execution. Numbered regions include prefrontal cortex, posterior middle temporal gyrus (pMTG), posterior superior temporal gyrus/superior temporal sulcus (pSTG/STS), Area Spt, inferior frontal gyrus/Broca's area (IFG), ventral premotor cortex (vPMC), supplementary and pre-supplementary motor areas (SMA/pre-SMA), and ventral sensorimotor cortex (vSMC). Arrows indicate primary production flow together with auditory and somatosensory/articulatory feedback pathways.
    \textbf{B.} Temporal progression through the speech hierarchy. Speech-related neural dynamics evolve from early acoustic--phonological analysis and auditory monitoring toward partially overlapping stages of auditory--motor interfacing, lexical access, phonological selection, sequencing, articulatory planning, and later contextual or conceptual stabilization. The schematic timeline emphasizes that these stages overlap rather than forming a strictly serial cascade.
    \textbf{C.} Conceptual latent spaces across the speech hierarchy. Different representational levels may support distinct decodable geometries in neural population activity: semantic/conceptual representations may separate categories such as animals versus tools or actions versus objects; phonological/auditory--motor representations may separate speech sounds such as /ba/ and /ga/; and motor representations in vSMC may separate articulatory classes such as labial versus coronal/dental gestures. These latent spaces are schematic rather than data-derived and are intended to illustrate how the choice of decoding target selects different levels of the speech hierarchy.}
    \label{fig:2}
\end{figure}

Ventral sensorimotor and premotor cortices encode articulatory gestures and motor plans \cite{Bouchard2013MotorCortexSpeech,Cheung2016ArticulatoryCortex}, whereas auditory cortex and superior temporal gyrus represent acoustic and phonetic structure, with increasing abstraction along temporal cortex \cite{Mesgarani2014PhoneticFeatures,Chang2010CategoricalSpeech, Goldsteinetal2025}. Inferior frontal and temporoparietal regions contribute to phonological sequencing, lexical access, and semantic integration \cite{Friederici2011LanguageNetwork,Blank2016SpeechHierarchy, Goldsteinetal2025}. Speech BCIs therefore sample different points in the hierarchy: reconstruction approaches often exploit auditory or sensorimotor representations \cite{Pasley2012Reconstruction,Akbari2019SpeechReconstruction}, while text decoders increasingly use higher-level linguistic structure \cite{moses2021neuroprosthesis,Willett2023SpeechBCI, card2024accurate, metzger2022generalizable}. These targets span articulatory and acoustic features, phonemes, syllables, words, and sentence- or meaning-level representations. They differ in abstraction, dimensionality, temporal granularity, stability, and task dependence, and may form distinct geometries in neural population activity. Choosing among them is therefore a substantive modelling decision rather than a simple change of output vocabulary. Figure~\ref{fig:2} summarizes the corresponding anatomical, temporal, and representational landscape.

\subsection{Sensorimotor integration and internal models}

Speech production is inherently sensorimotor. Neural systems responsible for generating speech movements are tightly coupled to those that process auditory and somatosensory feedback, forming internal models that predict the sensory consequences of motor commands \cite{Hickok2011DualStream,Guenther2016NeuralControlSpeech}. These predictive mechanisms enable rapid error correction and fluent speech, even in the presence of noise or perturbation.

Evidence from neurophysiology and neuroimaging indicates that motor regions encode not only outgoing articulatory commands but also predicted sensory outcomes \cite{Tourville2008AuditoryFeedback,Cheung2016ArticulatoryCortex}. Discrepancies between predicted and actual feedback generate error signals that drive adaptive updates to motor plans \cite{Houde1998SpeechPerturbation}. This tight coupling challenges simplistic interpretations of neural activity as purely motor or sensory and complicates the assignment of decoding targets in speech BCIs.

For BCIs, sensorimotor integration implies that neural signals used for decoding are shaped by both intended speech and the feedback context provided by the interface. Altering feedback modalities, latency, or accuracy can therefore change the underlying neural representations themselves, even when the overt task remains unchanged \cite{Ganguly2009Reorganization,Taylor2002LearningBCI}. This observation foreshadows the importance of closed-loop co-adaptation, discussed in later sections, and cautions against assuming a stationary mapping between neural activity and speech targets.

\subsection{Distributed versus localized representations}

Speech information is distributed across sensorimotor, temporal, and frontal populations, with redundancy and overlap across regions \cite{Chang2010CategoricalSpeech,Leonard2016DistributedSpeech}. Individual neurons and electrodes can multiplex several features depending on context \cite{Mesgarani2014PhoneticFeatures,Bouchard2013MotorCortexSpeech}. Pooling across populations can therefore improve robustness and compensate for local variability, as demonstrated with high-density ECoG and intracortical arrays \cite{herff2015brain,Willett2023SpeechBCI}, but distributed codes also complicate functional localization, electrode selection, and the transfer of fixed feature sets across users and sessions \cite{Sussillo2016NeuralDrift,perge2014reliability}. Spatial organization also depends on the level of the speech-processing hierarchy: articulatory and acoustic features often show more localized and structured cortical organization, whereas lexical and semantic information is represented more broadly across distributed cortical regions \cite{Huth2016SemanticMaps,Blank2016SpeechHierarchy}.

\subsection{Temporal dynamics, prediction, and feedback loops}

Neural activity during speech reflects planning, execution, monitoring, and correction on overlapping timescales \cite{Cogan2014TimeResolvedSpeech,Euston2012TimeScales}. Predictive-coding accounts propose that cortical systems continuously anticipate upcoming sensory input and motor states \cite{Friston2010PredictiveCodingSpeech}, motivating sequence models that capture temporal dependencies rather than decoding time points independently \cite{Sussillo2016RNN,glaser2020machine}. Speech BCIs also require alignment between continuous neural dynamics and discrete, hierarchically organized linguistic units whose boundaries may be variable, overlapping, or unobserved. Unlike continuous kinematic targets in many motor BCIs, phonemes, words, and sentences unfold at different rates and may lack directly observed onset times. Decoding must therefore jointly infer representation, segmentation, and sequence structure. Because delayed or inconsistent feedback can perturb predictive loops and degrade control \cite{Houde1998SpeechPerturbation,Ganguly2009Reorganization}, temporal alignment and system latency are fundamental design constraints.

\subsection{Implications for designing targets and labels}

Decoding targets implicitly select a representational level and assumptions about neural stability. Low-level targets may be vulnerable to noise and drift, whereas high-level targets may allow language priors to dominate neural evidence \cite{Akbari2019SpeechReconstruction,Willett2023SpeechBCI}. Target selection should therefore reflect the intended function of the interface, whether intelligible speech synthesis, rapid text communication, or naturalistic interaction, as well as the available neural coverage and supervision. Hybrid neural--linguistic models, self-supervised objectives, and explicit uncertainty estimation may improve transfer and robustness \cite{tang2023semantic,jayalath2024brain,Gal2016DropoutUncertainty}. More broadly, labels should be treated as design choices within an adaptive communication system rather than immutable ground truth. Their value depends not only on offline decodability, but also on interpretability, calibration burden, real-time controllability, and compatibility with user feedback.
\section{Recording modalities and hardware considerations}\label{modal_hardware}

Speech BCIs require recording interfaces that balance spatial resolution, temporal fidelity, cortical coverage, long-term stability, and clinical feasibility. No modality optimizes all of these dimensions. Surface systems sample population activity across broader cortical territories, penetrating arrays provide finer access to spiking and local field potentials, and endovascular interfaces reduce implantation burden at the cost of signal strength and spatial specificity. This diversity is reflected in Utah-style arrays \cite{Hochberg2012IntracorticalBCI,davidoff2020agency}, Neuralink's flexible-thread N1 \cite{musk2019integrated,Neuralink2024PRIME}, Synchron's Stentrode \cite{mitchell2023assessment}, and Precision Neuroscience's conformable Layer 7 surface interface \cite{hettick2025minimally}. Non-invasive methods avoid implantation but generally sacrifice some combination of spatial resolution, signal-to-noise ratio, or real-time practicality. Device geometry, channel count, materials, and implantation strategy therefore constrain both decodable information and longitudinal reliability \cite{davidoff2020agency,mitchell2023assessment,schroter2025advances}.

\subsection{Surface, depth, and endovascular intracranial recordings}

\paragraph{Electrocorticography (ECoG).}
ECoG combines millisecond-scale temporal resolution, relatively high signal quality, and broad cortical coverage with less tissue penetration than intracortical arrays. Standard subdural grids use millimetre-scale contacts, while newer high-density and thin-film systems improve sampling over speech-relevant cortex \cite{Leuthardt2004ECoG,fifer2013simultaneous}. ECoG is particularly sensitive to high-gamma activity (\SIrange{70}{150}{Hz}), a population-level signal associated with local firing and articulatory and phonetic representations \cite{Crone2001HighGamma,Bouchard2013MotorCortexSpeech}, and high-gamma features have supported phoneme decoding, word classification, and continuous speech reconstruction \cite{herff2015brain,Pasley2012Reconstruction,moses2021neuroprosthesis}. However, surface recordings do not resolve single neurons, primarily sample exposed gyral cortex, and are often placed according to clinical priorities. Coverage therefore varies across individuals, while electrode impedance, physiology, and behavioural context contribute to non-stationarity \cite{Sussillo2016NeuralDrift}.

\paragraph{Stereoelectroencephalography (sEEG).}
sEEG complements surface recordings by sampling deep and sulcal structures, including medial temporal, insular, and inferior frontal regions involved in speech and language \cite{Buzsaki2012SEEG}. Depth leads typically contain roughly 5--18 cylindrical contacts along three-dimensional trajectories \cite{li2018optimal,khoo2020technical}. These recordings contain phonetic and lexical information during speech perception and production \cite{Flinker2015SpeechSEEG,Chartier2018SEEGSpeech}, and permit investigation of hierarchical and cross-regional processing consistent with the distributed organization described in Section~\ref{neural_substrates}. Their sparse, individualized, clinically determined trajectories nevertheless limit standardization and cross-subject transfer.

\paragraph{Endovascular recording.}
Endovascular interfaces place electrodes within cerebral vessels near cortical targets, providing neural access without open-brain implantation. Their principal appeal is reduced procedural burden, although current systems trade this advantage for lower spatial specificity and signal amplitude.

\subsection{Intracortical microelectrode arrays}

Intracortical arrays provide the highest spatial and temporal resolution available in human BCIs by recording action potentials and local field potentials from small neuronal populations \cite{Hochberg2012IntracorticalBCI}. Classical Utah arrays, including NeuroPort, contain up to 96 penetrating silicon electrodes and underpin many landmark studies \cite{Hochberg2012IntracorticalBCI,davidoff2020agency}. Newer designs expand channel count and coverage; Neuralink's N1, for example, uses 64 flexible threads carrying 1,024 electrodes \cite{musk2019integrated}.

Motor-cortical recordings have enabled rapid word and sentence decoding at communication rates approaching natural speech under constrained conditions \cite{Willett2021Intracortical,Willett2023SpeechBCI}. This bandwidth comes with greater invasiveness and uncertain chronic reliability: tissue responses, micromotion, and electrode failures can degrade signal quality over months or years \cite{perge2014reliability,Barrese2016FailureModes}. Translation therefore depends not only on decoder performance but also on durable materials, surgical safety, and models that remain useful despite changing recorded units.

\subsection{Non-invasive modalities}

\paragraph{EEG, MEG, and fNIRS.}
EEG, MEG, and fNIRS avoid implantation and may suit users for whom surgery is inappropriate \cite{brumberg2010brain}. EEG and MEG offer high temporal resolution but limited spatial specificity and sensitivity to physiological and environmental artifacts \cite{Lalor2009EEGSpeech}; fNIRS provides somewhat better localization but its slow hemodynamic response is poorly suited to rapid speech control \cite{cooney2021bimodal}. MEG studies have demonstrated constrained word decoding and speech-segment retrieval, while naturalistic datasets such as MEG-MASC and LibriBrain have expanded opportunities for studying speech and language processing \cite{defossez2023decoding,gwilliams2023introducing,d2025towards,ozdogan2026libribrain}.
EEG language decoding has likewise expanded through datasets such as ZuCo, although open-vocabulary EEG-to-text remains disputed because apparent performance may partly reflect language-model priors, teacher forcing, memorization, or dataset artefacts \cite{hollenstein2018zuco,jo2025evaluating}. These modalities remain valuable for studying language representations and lower-bandwidth applications, but currently lag invasive systems for reliable, high-rate communication.

\paragraph{Functional MRI (fMRI).}
fMRI provides whole-brain spatial coverage but measures neural activity indirectly through slow hemodynamic responses. It is therefore unsuitable for low-latency speech prostheses, yet important for mapping distributed lexical, semantic, and discourse-level representations. Continuous natural language produces widespread semantic maps across temporal, parietal, and frontal cortex \cite{Huth2016SemanticMaps}. Large language models and encoding--decoding frameworks have extended this work to reconstruction of words, phrases, and narrative content from prolonged listening or reading \cite{antonello2024evidence,tang2023semantic}. These outputs primarily recover semantic gist and depend on long integration windows and strong contextual priors. fMRI consequently emphasizes higher-level representations, whereas invasive systems more often target sensorimotor or phonetic signals for rapid control (Sections~\ref{speech_rep} and~\ref{decoding}); abstract representations may be more distributed but are further from moment-to-moment intent and more vulnerable to linguistic priors (Section~\ref{speech_rep_mismatch}).

fMRI has also supported representational similarity analysis, semantic embedding alignment, self-supervised learning, and multimodal neural--language modelling \cite{Huth2016SemanticMaps,tang2023semantic}. These ideas increasingly influence electrophysiological speech BCIs and foundation-model approaches (Section~~\ref{decoding_ssl}), although prolonged data collection and strong model dependence can encourage overstatement of the fidelity with which internal speech or intent is recovered \cite{tang2023semantic}. This concern connects directly to mental privacy, agency, and the boundary between neural evidence and model inference (Sections~\ref{nat_comms} and~\ref{ethic_legal}). fMRI is therefore primarily a representational and methodological modality rather than a direct clinical communication interface.

\subsection{Datasets, preprocessing pipelines, and feature extraction standards}

Dataset construction and preprocessing often shape performance as strongly as model architecture. Invasive datasets generally arise from temporary presurgical ECoG/sEEG monitoring or dedicated chronic implants in paralysis and ALS \cite{herff2020potential,Willett2023SpeechBCI}. ECoG studies commonly include fewer than 15 participants and tens of minutes to several hours of recording \cite{moses2021neuroprosthesis,herff2015brain}, while intracortical studies involve very few implanted users and intensive within-user training \cite{Willett2023SpeechBCI}. By contrast, fMRI semantic decoding often requires hours of naturalistic exposure per participant \cite{Huth2016SemanticMaps,tang2023semantic}, exemplifying the ``deep neuroimaging paradigm'' \cite{Kupers2024}. Performance across these regimes is therefore not directly comparable.

ECoG pipelines commonly combine filtering, re-referencing, and standardized high-gamma envelopes \cite{Crone2001HighGamma}. Intracortical features may use sorted spikes or threshold crossings, with multiunit measures potentially more robust to chronic unit turnover \cite{perge2014reliability}. Non-invasive preprocessing emphasizes artifact suppression and weak-signal recovery: EEG commonly removes ocular and muscle artifacts and applies spatial or band-limited filtering \cite{Lalor2009EEGSpeech}, whereas fMRI requires motion correction, normalization, temporal filtering, and explicit modelling of hemodynamic delay \cite{Huth2016SemanticMaps}. Semantic decoding then estimates mappings between contextual word or sentence embeddings extracted from language models and BOLD responses for reconstruction \cite{tang2023semantic}.

Across modalities, neural time series must also be aligned to uncertain linguistic labels without leakage. Forced alignment can estimate phoneme or word timing, while continuous models often use context windows that include preparatory activity. Naturalistic speech increases boundary uncertainty, and fMRI adds temporal blurring. Because performance can vary with referencing, normalization, windowing, artifact rejection, alignment, and data splitting, reproducible studies should report participants, recording duration and coverage, preprocessing and features, label generation, alignment, and train/validation/test splits.

\subsection{Longitudinal stability and clinical constraints}

Long-term recordings change with electrode encapsulation, neural plasticity, cognitive state, and task context \cite{Sussillo2016NeuralDrift}. Representational drift can be substantial even when behaviour remains stable \cite{perge2014reliability}. Adaptive decoders, co-adaptive training, and robust representational targets may mitigate this variability \cite{Ganguly2009Reorganization,Degenhart2020Stability}, making adaptation a core requirement for chronic use.

Recording choices must also account for fatigue, cognitive load, comorbidities, and limited experimental tolerance in people with severe speech impairment \cite{felton2012mental,Wolpaw2018HomeUseNeurology}. Surgical risk, maintenance, and regulatory demands further constrain chronic systems, while lesion extent, neural reorganization, and compensatory strategies limit straightforward model transfer \cite{felton2012mental}. Hardware and decoding strategies must therefore be matched to realistic communication goals and designed for sustained, individualized adaptation.

In the next section, we examine how choices of speech representation and decoding target interact with these recording constraints.

\section{Speech representations: what can (and should) be decoded?}\label{speech_rep}

The choice of representation determines what information a speech BCI must extract, which errors it can tolerate, and how strongly it depends on downstream priors. As discussed in Sections~\ref{neural_substrates} and~\ref{modal_hardware}, neural recordings provide partial access to a hierarchical and dynamic speech system. Articulatory, acoustic, phonological, lexical, semantic, and pragmatic targets therefore differ in their neural accessibility, temporal granularity, interpretability, generalization, and clinical utility.

\begin{itemize}
    \item \textbf{Articulatory representations}

Articulatory representations describe configurations and movements of the lips, tongue, jaw, and larynx. They align closely with ventral sensorimotor and premotor activity, making them natural targets for invasive BCIs \cite{Bouchard2013MotorCortexSpeech,Cheung2016ArticulatoryCortex}. ECoG and intracortical studies have inferred articulatory kinematics and converted them to intelligible speech through articulatory-to-acoustic synthesis \cite{Chartier2018ArticulatoryBCI,Anumanchipalli2019SpeechSynthesis}. These targets can be interpretable and less speaker-specific than raw acoustics, but require assumptions and auxiliary models linking neural activity, kinematics, and sound, and may be less applicable to imagined speech without overt execution \cite{martin2018decoding}.

    \item \textbf{Acoustic representations}

Acoustic targets, including spectrograms and mel-frequency features, support direct reconstruction of audible speech from neural activity \cite{Pasley2012Reconstruction,Akbari2019SpeechReconstruction}. Their physical structure relates to activity in auditory and sensorimotor cortex \cite{Mesgarani2014PhoneticFeatures}, and neural vocoders can transform decoded features into intelligible, natural-sounding output \cite{Anumanchipalli2019SpeechSynthesis}. However, acoustic decoding is sensitive to timing errors and noise, often requires substantial post-processing, and can entangle intended production with auditory feedback \cite{Tourville2008AuditoryFeedback}, limiting interpretation, generalization, and closed-loop control.

    \item \textbf{Phonetic and phonological units}

Phonetic and phonological units provide an intermediate representation between gestures and words. Superior temporal and sensorimotor populations encode features such as place and manner of articulation \cite{Mesgarani2014PhoneticFeatures,Chang2010CategoricalSpeech}, motivating phoneme and phoneme-sequence decoders \cite{herff2015brain,Mugler2014PhonemeBCI}. Phonemes reduce output dimensionality and integrate naturally with language models \cite{Moses2019SentenceDecoding}, but externally defined inventories may not match neural organization, differ across languages and speakers, and require uncertain temporal alignment. Errors can also propagate when noisy phoneme sequences are converted into words.

    \item 
\textbf{Lexical, semantic, and pragmatic representations}

Word-, sentence-, and semantic-level decoders bypass lower-level representations and can support rapid text communication under constrained conditions \cite{moses2021neuroprosthesis,Willett2023SpeechBCI}. Such targets may be less sensitive to acoustic variability and potentially more stable across sessions \cite{Huth2016SemanticMaps}. Their central risk is that strong language models can produce fluent output weakly constrained by neural evidence, obscuring user intent and agency \cite{tang2023semantic}. Pragmatic information—including conversational goals, context, and intended social action—remains largely unexplored despite being essential to natural communication and will require richer models of user state and interaction.

\end{itemize}

\phantomsection
\label{speech_rep_prosody}

Beyond these primary representational targets, prosody conveys emphasis, emotion, uncertainty, turn structure, and speaker identity through intonation, rhythm, stress, rate, and intensity. Its correlates in auditory and motor regions suggest that these features are decodable in principle \cite{tang2017intonational,Cogan2014TimeResolvedSpeech}. Yet most BCIs optimize lexical accuracy and intelligibility, leaving output flattened or monotonic even when the words are correct. Prosody is difficult to model because it is variable, subtle, and distributed across timescales, and conventional metrics such as word error rate do not measure its preservation. Naturalistic systems will therefore need to treat affect and prosody as both decoding targets and evaluation dimensions.

\phantomsection
\label{speech_rep_modes}

In addition to what is decoded, representational accessibility also depends on how speech is produced. Overt speech provides strong motor and sensory signals but is unavailable to many intended users; mouthed speech retains articulation without sound and has been decoded successfully \cite{herff2015brain}. Imagined speech is clinically attractive because it requires no overt movement, but its neural correlates are weaker, more variable, and not necessarily equivalent to overt articulatory or acoustic representations \cite{martin2018decoding,cooney2021bimodal}. Robust decoding may therefore require distinct targets and training paradigms rather than straightforward transfer from overt speech.

\phantomsection
\label{speech_rep_mismatch}

Across both representational targets and behavioural modes, externally imposed labels may not align with the representations encoded by neural populations \cite{Hickok2012SpeechNeuralBasis}. This mismatch can produce brittle models, uncertain alignment, and poor transfer even when within-dataset accuracy is high. Adaptive targets, self-supervised objectives, and hybrid models may instead allow useful representations to emerge from paired or unlabelled data \cite{defossez2023decoding,jayalath2024brain}. What should be decoded is therefore inseparable from how the system is trained, evaluated, and used: the optimal representation is not simply the most decodable offline, but the one that best supports accurate, controllable, and generalizable communication.

In the next section, we examine how decoding paradigms and model families operationalize these representational choices.

\section{Decoding paradigms and model families}\label{decoding}

The choice of decoding paradigm determines not only how neural activity is translated into speech-related outputs, but also what assumptions are made about temporal alignment, representational level, supervision, and uncertainty. Sections~\ref{neural_substrates}--\ref{speech_rep} emphasized that speech-related neural activity is hierarchical, distributed, and dynamic, and that recording modalities provide incomplete and non-stationary access to this space. Decoders therefore operate under fundamental constraints: they must map high-dimensional, noisy, and drifting neural signals onto speech representations whose appropriate level may itself be uncertain.

\begin{figure}[!h]
    \centering
    \includegraphics[width=0.9\linewidth]{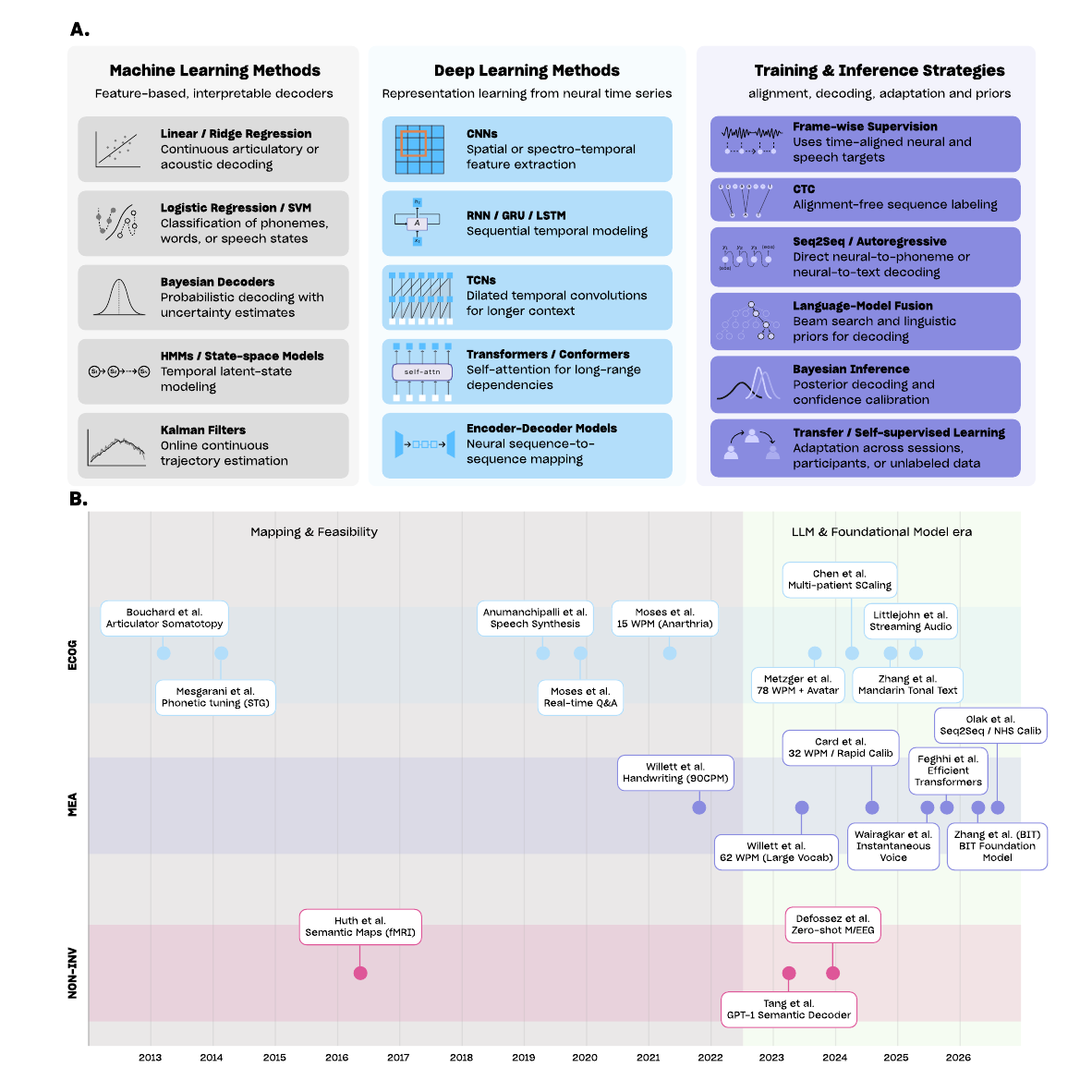}
    \caption{\captiontitle{Methods and evolution of speech BCI decoding.}
    \textbf{A.} Taxonomy of decoding models and training strategies. Speech BCI pipelines span three complementary levels. Machine-learning and probabilistic approaches provide feature-based mappings from neural activity to speech-related targets, including regression and classification models, Bayesian decoders, HMMs and state-space models, and Kalman filters. Deep learning approaches learn representations directly from neural time series using CNNs, recurrent and temporal convolutional networks, transformers or conformers, and encoder--decoder architectures. Training and inference strategies determine how neural activity is aligned with speech outputs and how prior information is incorporated, including frame-wise supervision, CTC, sequence-to-sequence or autoregressive decoding, language-model fusion, Bayesian inference, and transfer or self-supervised learning. Importantly, these strategies are distinct from the underlying model architecture; for example, CTC and language-model fusion specify alignment or inference mechanisms rather than decoder architectures.
\textbf{B.} Evolving landscape of speech BCI research. Representative studies are positioned by approximate publication year and recording modality, including electrocorticography (ECoG), intracortical microelectrode-array systems (MEA), and non-invasive approaches such as fMRI, EEG, and MEG. The timeline highlights major methodological and decoding milestones, from articulatory, phonetic, acoustic, and semantic mapping to speech synthesis, large-vocabulary text decoding, rapid calibration, streaming speech or audio generation, and foundation-model-based approaches. Background shading distinguishes an earlier \emph{mapping and feasibility} period from the more recent \emph{LLM and foundation-model era}, reflecting a broader shift toward high-throughput, clinically oriented, scalable, and language-aware communication systems. The timeline is selective rather than exhaustive and summarizes major trends across the field.}
    \label{fig:3}
\end{figure}

Figure~\ref{fig:3}.A summarizes speech BCI decoding along three complementary axes: classical machine-learning and probabilistic decoders, deep learning architectures, and training or inference strategies that address alignment, uncertainty, adaptation, and linguistic priors.

At its most general, speech decoding can be framed as conditional sequence modeling:
\[
\mathbf{y}_{1:T} \sim p_\theta(\mathbf{y}_{1:T} \mid \mathbf{x}_{1:U}),
\]

where $\mathbf{x}_{1:U}$ denotes a multichannel neural time series and $\mathbf{y}_{1:T}$ denotes an output sequence (articulatory trajectories, acoustic features, phonemes, characters, or words). The parameters $\theta$ define the decoder architecture. Different model families correspond to different structural assumptions about this conditional distribution — including how time is modeled, how alignment is handled, and how uncertainty is represented.

\subsection{Classical pipelines and linear decoders}

Early BCIs often relied on modular pipelines consisting of neural feature extraction followed by a linear mapping to a behavioural or speech target \cite{Crone2001HighGamma,Leuthardt2004ECoG}. Linear models remain particularly effective for relatively low-dimensional motor-control tasks, such as decoding reaching kinematics, where neural population activity can often be mapped reliably to continuous movement variables. In speech BCIs, linear, regularized linear, and logistic regression remain widely used as strong and interpretable baselines \cite{herff2015brain,Mugler2014PhonemeBCI}.

In its simplest form, regularized linear regression decoding can be written as
\[
\hat{\mathbf{y}} = W \mathbf{x} + b,
\]
with parameters estimated by minimizing
\[
\mathcal{L}_{\text{MSE}} = \|\mathbf{y} - W\mathbf{x}\|_2^2 + \lambda \|W\|_2^2.
\]

Here, $\mathbf{x} \in \mathbb{R}^d$ denotes a neural feature vector extracted from a time window (e.g., high-gamma power across electrodes), $\mathbf{y} \in \mathbb{R}^k$ denotes the corresponding speech target (e.g., acoustic features or one-hot phoneme labels), and $\hat{\mathbf{y}}$ is the model prediction. In practice, $\mathbf{x}$ is often constructed from multichannel neural activity over a fixed time window, implicitly assuming a predefined alignment between neural activity and the corresponding speech target.

The appeal of this formulation lies in its bias--variance balance. With limited data — common in invasive BCI studies — strong regularization can yield stable generalization. Moreover, high-gamma power correlates approximately linearly with local firing rates \cite{Crone2001HighGamma,Bouchard2013MotorCortexSpeech}, making linear models surprisingly competitive within-session.

However, linear mappings cannot capture nonlinear population codes, long-range dependencies, or hierarchical linguistic structure. They also assume fixed alignment between neural features and output labels, an assumption often violated in continuous speech tasks.

\subsection{State-space and probabilistic models}

Speech unfolds dynamically, and neural signals often reflect latent articulatory or planning processes. State-space models introduce explicit latent dynamics:
\begin{align}
\mathbf{z}_t &= A\mathbf{z}_{t-1} + \mathbf{w}_t, \\
\mathbf{x}_t &= C\mathbf{z}_t + \mathbf{v}_t,
\end{align}
where $\mathbf{z}_t$ represents latent speech states, and $\mathbf{w}_t \sim \mathcal{N}(0,Q)$ and $\mathbf{v}_t \sim \mathcal{N}(0,R)$ are typically modeled as zero-mean Gaussian process and observation noise, respectively.

Inference computes
\[
p(\mathbf{z}_t \mid \mathbf{x}_{1:t}),
\]
which, in decoding settings, is obtained by inverting this generative model to estimate latent speech states from observed neural activity, typically via Kalman filtering \cite{Wu2006KalmanBCI,Shenoy2013StateSpaceBCI}

The advantage is principled temporal smoothing and uncertainty propagation. The limitation is that linear-Gaussian assumptions may mismatch nonlinear cortical dynamics and complex phonetic transitions.

Hidden Markov models (HMMs) extend this idea to discrete sequences \cite{Rabiner1989HMM}, modeling phoneme transitions probabilistically. These models encode discrete latent states and probabilistic transitions, enabling explicit modeling of phoneme-level alignment via dynamic programming, but require carefully designed state structures and sufficient labeled data.

\subsection{Deep neural architectures: CNNs, RNNs, and TCNs}

Deep learning reduces reliance on hand-engineered features by learning hierarchical representations directly from neural signals. Convolutional neural networks (CNNs) exploit local spatiotemporal structure across electrode grids \cite{Schirrmeister2017DeepEEG,Lawhern2018EEGNet}. 

Recurrent neural networks (RNNs) model temporal dependencies via hidden states:

\[
\mathbf{h}_t = f_\theta(\mathbf{x}_t, \mathbf{h}_{t-1}).
\]

Here, $f_\theta$ denotes a nonlinear transition function parameterized by $\theta$, and $\mathbf{h}_t$ represents the hidden state summarizing past neural activity.

Vanilla RNNs suffer from vanishing gradients. Gated architectures such as LSTMs \cite{Hochreiter1997LSTM} and GRUs introduce update mechanisms:

\[
\mathbf{h}_t = (1 - \mathbf{z}_t)\odot \mathbf{h}_{t-1} + \mathbf{z}_t \odot \tilde{\mathbf{h}}_t.
\]

Here, $\mathbf{z}_t$ denotes a learned update gate (not to be confused with the latent state in Section~\ref{decoding}), $\tilde{\mathbf{h}}_t$ denotes the candidate hidden state proposed from the current input and previous hidden state, and $\odot$ denotes element-wise multiplication.

Intuitively, gates allow the network to retain long-range information while filtering noise. GRUs are often preferred in BCI settings due to fewer parameters and greater data efficiency, whereas LSTMs may provide greater flexibility at the cost of increased sample complexity.

Temporal convolutional networks (TCNs) use dilated causal convolutions to achieve large receptive fields without recurrence \cite{Bai2018TCN}, enabling stable gradient propagation and low-latency streaming inference—properties particularly valuable for real-time BCIs.

\subsection{Alignment and sequence objectives: CTC and encoder--decoder models}

A central challenge in neural-to-text decoding is unknown temporal alignment between neural activity and linguistic units. Connectionist Temporal Classification (CTC) addresses this by marginalizing over monotonic alignments \cite{graves2006connectionist}:

\[
\mathcal{L}_{\text{CTC}} = - \log \sum_{\pi \in \mathcal{A}(\mathbf{y})} p_\theta(\pi \mid \mathbf{x}),
\]

where $\pi$ denotes a path (frame-level labeling including blank symbols), $\mathbf{x}$ represents the input neural sequence, and $\mathcal{A}(\mathbf{y})$ denotes valid alignments.

CTC avoids requiring frame-level phoneme labels and supports streaming inference. However, CTC assumes conditional independence between output symbols given the encoder representations, which can limit modeling of long-range linguistic dependencies and often motivates integration with external language models.

Encoder--decoder models instead directly optimize:

\[
\mathcal{L}_{\text{CE}} = - \sum_{t=1}^{T} \log p_\theta(y_t \mid y_{<t}, \mathbf{x}),
\]

which corresponds to an autoregressive factorization of the output sequence, where each token is conditioned on previously generated outputs. This allows output tokens to condition on prior outputs, improving linguistic coherence but increasing reliance on learned priors, intensifying the need for calibration and ablation analysis.

\subsection{Transformers and attention mechanisms}

Transformers replace recurrence with self-attention:

\[
\text{Attention}(Q,K,V) = \text{softmax}\!\left(\frac{QK^\top}{\sqrt{d_k}}\right)V
\]

Here, $Q$, $K$, and $V$ denote learned query, key, and value projections computed from the input neural sequence $\mathbf{x}$, and $d_k$ is the key dimensionality.

Self-attention enables flexible integration over long temporal windows and selective weighting of informative neural segments \cite{Vaswani2017Attention}. In speech BCIs, this can improve robustness to local noise and facilitate sequence modeling.

However, high-capacity attention models risk overfitting in small datasets, and when combined with strong external language priors (e.g., via rescoring or constrained decoding), they can blur the boundary between neural evidence and model-driven inference.

\subsection{Self-supervised and contrastive objectives}\label{decoding_ssl}

Limited labeled data has motivated self-supervised learning (SSL), including contrastive, reconstruction-based, and predictive objectives. These approaches can exploit large unlabelled neural datasets to learn more transferable representations, or leverage representations learned from data-rich modalities such as speech audio to provide useful structure for neural decoding \cite{zhangcross}. Contrastive objectives such as InfoNCE encourage alignment between neural and speech embeddings:

\[
\mathcal{L}_{\text{InfoNCE}} = - \log \frac{\exp(\mathbf{z}_i \cdot \mathbf{z}_i^+ / \tau)}{\sum_j \exp(\mathbf{z}_i \cdot \mathbf{z}_j / \tau)}.
\]

Here, $\mathbf{z}_i$ denotes an anchor embedding (e.g., neural), $\mathbf{z}_i^{+}$ its matched positive (e.g., paired speech), $\{\mathbf{z}_j\}$ are candidate embeddings (including negatives), and $\tau>0$ is a temperature parameter.

Geometrically, matched neural--speech pairs are pulled together in embedding space, while mismatched pairs are pushed apart. However, contrastive learning depends critically on defining meaningful positive and negative pairs. This is difficult when neural activity and speech targets are not explicitly aligned, as in continuous, imagined, or naturalistic speech paradigms. In such cases, models may learn shortcuts based on stimulus timing, task structure, session identity, or other non-speech covariates rather than communicative intent.

Other SSL objectives may partly mitigate this problem. Reconstruction losses train models to recover masked or corrupted neural or speech features from context, whereas predictive approaches such as joint embedding predictive architectures (JEPAs) aim to predict abstract target embeddings rather than reconstructing low-level inputs \cite{lecun2022path,assran2023self}. These objectives may improve data efficiency and cross-session robustness, but their success depends on designing pretext tasks that capture speech-relevant neural structure rather than artefacts of the experimental protocol.

\subsection{Multimodal alignment and foundation models}

Multimodal models connect neural activity with acoustic, articulatory, speech-model, or language-model representations. Contrastive alignment between MEG or EEG and self-supervised speech embeddings has enabled retrieval of perceived speech segments \cite{defossez2023decoding}. In invasive BCIs, related systems first infer articulatory or acoustic representations and then synthesize speech or text \cite{Anumanchipalli2019SpeechSynthesis,chen2024neural,metzger2023high}.

Foundation models contribute pretrained acoustic spaces and vocoders, language priors for large-vocabulary decoding, and transferable neural representations. Language models can correct noisy phoneme or subword sequences \cite{moses2021neuroprosthesis,Willett2023SpeechBCI,tang2023semantic}, but fluent output may exceed what is supported by the neural signal. Emerging neural foundation models instead seek transfer across subjects, tasks, and recording modalities through large-scale self-supervised pretraining \cite{zhang2025decoding}. Their value depends on genuine transfer, robustness to distribution shift, and separation of neural evidence from pretrained priors.

\subsection{Uncertainty estimation and calibration}\label{decoding_uncertainty}

Reliable clinical deployment requires calibrated uncertainty. Predictive variance can be decomposed as:

\[
\text{Var}(y \mid x) = \mathbb{E}_{\theta}[\text{Var}(y \mid x, \theta)] 
+ \text{Var}_{\theta}[\mathbb{E}(y \mid x, \theta)].
\]

Here, the expectations are taken with respect to uncertainty over model parameters $\theta$ (e.g., an approximate posterior), separating aleatoric uncertainty $\mathbb{E}_{\theta}[\mathrm{Var}(y\mid x,\theta)]$ from epistemic uncertainty $\mathrm{Var}_{\theta}[\mathbb{E}(y\mid x,\theta)]$.

Monte Carlo dropout approximates Bayesian model averaging by sampling different dropout masks at inference:
\[
\hat{p}(y \mid x) \approx \frac{1}{M} \sum_{m=1}^{M} p_{\theta}^{(m)}(y \mid x),
\]
where $p_{\theta}^{(m)}$ denotes the predictive distribution under the $m$-th dropout sample.

Well-calibrated uncertainty enables abstention, fallback strategies, and user-in-the-loop correction. This is particularly important when language models amplify fluency despite weak neural evidence.

\vspace{0.5em}
Across model families, architectural sophistication alone does not guarantee meaningful decoding. Objective functions encode assumptions about alignment and independence; model capacity shapes reliance on priors; supervision regime determines generalization limits. As speech BCIs evolve toward interactive, real-world systems, decoding must be understood not merely as function approximation, but as probabilistic inference under uncertainty within a closed-loop adaptive system. The next section therefore turns to closed-loop learning and co-adaptation, outlining why interactive dynamics are central to stability, generalization, and clinical viability.

\section{Empirical landscape of speech BCI studies: paradigms, modalities, and models}\label{empirical_landscape}

Encoding studies identify which speech features are represented in neural activity, where they occur, and over what timescales; decoding studies ask whether those signals can be transformed into useful outputs such as phonemes, words, text, synthesized speech, or communication actions. The two are complementary: encoding identifies candidate regions, representations, and temporal windows, whereas decoding tests whether they are sufficiently robust for communication. Their progression across modalities is summarized in Figure~\ref{fig:3}.B.

This distinction matters because speech BCIs target multiple representational levels. Ventral sensorimotor cortex contains articulatory structure relevant to production, whereas superior temporal cortex carries phonetic and acoustic information relevant to perception \cite{Bouchard2013MotorCortexSpeech,Mesgarani2014PhoneticFeatures}; distributed fMRI responses can additionally support semantic reconstruction when low-level articulatory detail is inaccessible \cite{tang2023semantic}. Intracranial studies further show that speech information is distributed across motor, premotor, temporal, and association regions \cite{kellis2010decoding,Bouchard2013MotorCortexSpeech,conant2018human,jamali2024semantic}. Recording site, target representation, and model family should therefore be considered jointly: implantation strategy is effectively a hypothesis about which level of the speech hierarchy can best support the intended communication goal.

\subsection{Empirical landscape across recording modalities}

\paragraph{Intracortical and MEA studies.}
Intracortical and microelectrode-array (MEA) systems provide temporally precise, fine-grained access to motor-cortical population activity. Early studies established that these signals contain phoneme- and word-related information in people with paralysis \cite{stavisky2019neural,wilson2020decoding}, building on BrainGate demonstrations that useful neural ensembles can be recorded over years \cite{hochberg2006neuronal,simeral2011neural}. Their principal advantage is high-bandwidth, low-latency access to attempted motor speech, although stability and calibration remain central constraints.

The empirical trajectory has moved from structured motor communication toward direct speech. Attempted handwriting showed that high-rate communication can be restored through any expressive and reliably decodable motor sequence, not only speech itself \cite{Willett2021Intracortical}. Subsequent systems decoded attempted phonemes from ventral precentral cortex and combined them with language models for large-vocabulary text output \cite{Willett2023SpeechBCI}. Rapid calibration and sustained performance in ALS further moved intracortical speech BCIs toward practical deployment \cite{card2024accurate}. In particular, large-vocabulary accuracy with only brief daily calibration \cite{card2024accurate} suggests that the bottleneck is shifting from signal availability toward longitudinal stability. Across studies, the strongest results generally combine low-level motor or subword targets, sequential neural encoders, and language priors rather than unconstrained end-to-end sentence generation.

\medskip
\paragraph{ECoG and sEEG studies.}
ECoG has generated the broadest speech BCI literature because it balances cortical coverage, temporal resolution, and signal quality over ventral sensorimotor and temporal regions. Foundational mapping studies established distributed articulatory organization in sensorimotor cortex and phonetic feature coding in superior temporal gyrus \cite{Bouchard2013MotorCortexSpeech,Mesgarani2014PhoneticFeatures}. Early decoders then classified phonemes or mapped phone-level activity to words \cite{Mugler2014PhonemeBCI,herff2015brain}, establishing the value of intermediate speech units and structured language models.

ECoG systems subsequently expanded from text to speech synthesis and multimodal output. Articulatory intermediate representations improved cortical speech synthesis \cite{Anumanchipalli2019SpeechSynthesis}; attempted-speech decoding enabled clinically useful communication in anarthria \cite{moses2021neuroprosthesis}; and later systems supported silent spelling \cite{metzger2022generalizable} as well as integrated text, personalized audio, and avatar control \cite{metzger2023high}. These studies position ECoG as a versatile substrate for restoring more than lexical content.

More recent systems target continuous, low-latency, and expressive output, combining text, streaming audio, facial animation, and conversational timing \cite{metzger2023high,Littlejohn2025StreamingBrainToVoice,wairagkar2025instantaneous}, while emphasizing prosody and vocal identity as clinically meaningful dimensions \cite{wairagkar2025instantaneous}. High-fidelity vocoders have begun to recover person-specific tone and prosodic characteristics \cite{wairagkar2025instantaneous} alongside multimodal communication \cite{metzger2023high}. Performance comparisons should therefore extend beyond word error rate to latency, intelligibility, expressivity, and preservation of social identity. Depth and mixed intracranial recordings provide complementary access to silent and imagined speech, including low- and cross-frequency features that may differ from overt articulatory signals \cite{proix2022imagined}; although sEEG has produced fewer high-throughput prostheses, it remains valuable for sampling distributed and deeper language networks.

\medskip
\paragraph{Non-invasive and peripheral approaches.}
Non-invasive methods occupy a different empirical niche. Large-scale contrastive learning has enabled retrieval of perceived speech representations from MEG and EEG \cite{defossez2023decoding}, while earlier MEG studies found decodable information in imagined and spoken phrases at substantially lower performance than invasive systems \cite{dash2020decoding}. fMRI is unsuitable for real-time prosthetic control but has demonstrated reconstruction of semantic and narrative content through language-model-based encoding and decoding \cite{tang2023semantic}, shifting attention toward distributed high-level representations and contextual priors.

Peripheral silent-speech systems provide an informative boundary case. EMG decodes articulatory muscle activity rather than cortical signals, yet shares problems of sequence alignment, speaker variability, and absent acoustic output \cite{denby2010silent}. Surface EMG can support recognition and synthesis when residual muscle activity remains \cite{vojtech2021surface}, making it a useful comparator for determining when invasive access provides clinically meaningful gains. Semantic fMRI mapping further shows that word meanings occupy reproducible distributed cortical spaces \cite{Huth2016SemanticMaps}; although not a direct communication technology, these maps inform hypotheses about where higher-level intent may be accessible.

\subsection{Evolution of communication paradigms}

Across modalities, communication BCIs have progressed from indirect selection, through structured motor-sequence decoding, toward direct speech restoration. Early P300 spelling used discrete, time-locked choices that were straightforward to train and evaluate \cite{Farwell1988P300,kubler2009brain}; intracortical cursor systems improved throughput but retained point-and-click interaction \cite{pandarinath2017high}. Attempted handwriting then demonstrated that an expressive motor alphabet could support rapid communication without reconstructing speech itself \cite{Willett2021Intracortical}.

Direct speech is more naturalistic but introduces variable timing, larger output spaces, and stronger dependence on linguistic priors. Imagined speech is additionally uncertain because the engaged production stages and available signals differ across individuals \cite{proix2022imagined}. High-performing clinical systems have therefore concentrated on attempted speech in anarthria or severe dysarthria, where articulatory intent may persist despite absent acoustic output \cite{moses2021neuroprosthesis,Willett2023SpeechBCI,card2024accurate}. The appropriate paradigm should therefore be selected according to the user's residual neural and motor capacities rather than a universal hierarchy of interfaces.

\subsection{From laboratory performance to user-centred utility}

Three trends complicate direct comparison across the empirical literature. First, sensor design is diversifying from penetrating arrays toward surface, depth, endovascular, and minimally invasive systems, making the relevant question not simply which decoder performs best, but which interface best matches the communication target, surgical constraints, and desired longevity. Second, effectors now span cursor control, typing, handwriting, text, synthesized voice, and avatars, each imposing different requirements for latency, vocabulary, calibration, and effort. Third, studies vary substantially in recording sites, channel counts, vocabularies, preprocessing, and predominantly single-participant protocols. An apparent gain may therefore reflect more favourable cortical coverage, a narrower task, additional training data, or stronger linguistic post-processing rather than a superior neural decoder.

Comparisons should accordingly report anatomical coverage, recording duration, vocabulary and behavioural paradigm, alignment method, calibration schedule, and the contribution of downstream priors. Neural foundation models offer one possible route toward shared representations by pretraining on larger non-invasive datasets and adapting to scarce invasive data \cite{defossez2023decoding}. Lightweight test-time adaptation and temporally masked architectures likewise aim to maintain performance across days while reducing computation and recalibration \cite{feghhi2025time}. Whether these approaches genuinely improve transfer requires standardized reporting, participant-independent tests where feasible, and leakage-resistant benchmarks.

These concerns become more important as evaluation shifts from peak laboratory accuracy toward reliability, independence, and sustainable use. Early home-use studies showed that long-term value depends on maintenance, electrode stability, and operation outside researcher-controlled settings \cite{vansteensel2016fully,vansteensel2024longevity,oxley2021motor,mitchell2023assessment}. More recent work emphasizes digital participation and clinically meaningful functional outcomes \cite{fry2022evaluating,sawyer2024digital,brannigan2024brain,dohle2025toward}, yet standardized patient-centred measures remain limited \cite{dohle2025toward}. Word error rate and communication speed remain necessary, but do not capture whether users can initiate interaction independently, recover from errors, communicate across partners and contexts, or sustain use without excessive researcher or caregiver support. Clinical utility therefore depends on minimizing calibration, fatigue, cognitive load, and correction cost while considering the complete system---including setup, feedback, user interface, technical support, and fallback modes.

\subsection{From neural decoding to ASR-inspired pipelines}

Alongside these shifts, contemporary neural speech decoders increasingly resemble automatic speech recognition (ASR) systems: neural activity is mapped to phonemes, characters, articulatory states, or acoustic embeddings, which are then converted to text or synthesized speech through language models and vocoders \cite{herff2016automatic,Anumanchipalli2019SpeechSynthesis,moses2021neuroprosthesis,Willett2023SpeechBCI}. CTC handles unknown monotonic alignments between continuous neural streams and symbol sequences \cite{graves2006connectionist}; recurrent transducers and sequence-to-sequence models support streaming and broader temporal context \cite{Graves2012RNNTransducer,chan2016listen,olak2026decoding}. Cross-task and cross-species pretraining extends this approach by aligning attempted and imagined speech with audio-language representations \cite{zhangcross}.

The analogy is nevertheless incomplete. Neural datasets are far smaller than acoustic corpora, recordings drift over time, and the mapping from neural activity to linguistic intent is less direct than that from sound to text. Neural systems therefore require modality-specific inductive biases, self-supervised objectives, and multi-task supervision \cite{chen2024neural}. Streaming also imposes causal constraints often hidden by offline decoding, while strong language models can conceal neural errors through plausible continuations. Systems therefore require ablations separating neural evidence from gains due to linguistic priors, alongside tests of latency, calibration, and stability. Without such controls, laboratory and participant heterogeneity makes it difficult to attribute improvements to modelling rather than recording conditions or post-processing.

Collectively, the literature supports three conclusions. Speech representations are distributed across cortical levels \cite{Bouchard2013MotorCortexSpeech,Mesgarani2014PhoneticFeatures,tang2023semantic}; intermediate targets such as phonemes, articulatory gestures, and handwriting strokes are often robust in data-limited settings \cite{Mugler2014PhonemeBCI,Willett2021Intracortical,Willett2023SpeechBCI}; and invasive systems currently provide the highest-throughput communication, whereas non-invasive methods remain important for perceptual and semantic modelling \cite{moses2021neuroprosthesis,defossez2023decoding,tang2023semantic}. Language priors are now integral to performance, but their contribution must be distinguished from neural information. Table \ref{tab:speech_bci_summary} summarizes representative studies, methods, and outcomes. The central challenge is no longer demonstrating that speech-related information is decodable, but matching interfaces, representations, and models to the clinical and communicative needs of individual users.



\begin{table*}[!ht]

\centering

\caption{Comprehensive Empirical Landscape: Anatomical Targets, Modalities, and Data Granularity}
\label{tab:speech_bci_summary}

\begin{tcolorbox}[
    colback=tether-gray,
    colframe=tether-gray,
    boxrule=0pt,
    arc=5pt,
    left=5pt,
    right=5pt,
    top=10pt,
    bottom=4pt,
    width=\textwidth
]

\fontsize{6.7pt}{6.2pt}\selectfont
\renewcommand{\arraystretch}{1.02}
\setlength{\tabcolsep}{2pt}
\setlength{\parskip}{0pt}
\setlength{\parindent}{0pt}

\begin{tabularx}{\linewidth}{
    >{\hsize=0.55\hsize\linewidth=\hsize\raggedright\arraybackslash}X
    >{\hsize=0.75\hsize\linewidth=\hsize\raggedright\arraybackslash}X
    c
    >{\hsize=0.70\hsize\linewidth=\hsize\raggedright\arraybackslash}X
    >{\hsize=0.70\hsize\linewidth=\hsize\raggedright\arraybackslash}X
    >{\hsize=0.90\hsize\linewidth=\hsize\raggedright\arraybackslash}X
    >{\hsize=2.40\hsize\linewidth=\hsize\raggedright\arraybackslash}X
}

\toprule

\textbf{Reference} &
\textbf{Modality / Site} &
\textbf{N} &
\textbf{Task Type} &
\textbf{Target} &
\textbf{Data Scope} &
\textbf{Method \& Detailed Outcomes} \\

\midrule
\rowcolor{rowblue}
\multicolumn{7}{l}{
\rule[0.4ex]{0pt}{3.2ex}
\raisebox{0.3ex}{\textbf{ECoG and Surface Arrays}}
} \\[0.4ex]
\cite{Moses2019SentenceDecoding} & ECoG (vSMC/STG) & 3 & Q\&A (Overt) & Utterances & 256 ch | Acute | 9 Q / 24 A & \textbf{HMM + Viterbi + Context Integration:} Real-time dialogue decoding. Max accuracy: 76\% (perceived Q) and 61\% (produced A); chance levels: 20\% and 7\% respectively. \\

\cite{moses2021neuroprosthesis} & ECoG (vSMC) & 1 (Anarthria) & Attempted & Words / Sentences & 128 ch | 48 sess (22h) | 50 words & \textbf{Deep Learning + LM:} Real-time sentence decoding; 25.6\% median WER at 15.2 WPM. Post hoc word classification accuracy: 47.1\%. \\

\cite{Bouchard2013MotorCortexSpeech} & ECoG (vSMC) & 3 & Overt (CV Syllables) & Articulator / Phonetic Features & $\sim$30 active ch/subj | 15--100 trials / syllable | 19 Cons. + 3 Vow. & \textbf{GLM + State-space PCA:} Demonstrated somatotopic layout of articulators (Larynx-Lips-Jaw-Tongue) and emergence of a phonetic feature hierarchy. \\

\cite{Mesgarani2014PhoneticFeatures} & ECoG (STG) & 6 & Perception (Continuous) & Phonetic Features & 256 ch | 500 sentences | English Phonemes & \textbf{STRF + PSI:} Found local selectivity to phonetic features (manner $>$ place). Demonstrated a distributed acoustic-phonetic representation in STG. \\

\cite{Anumanchipalli2019SpeechSynthesis} & ECoG (vSMC / STG) & 5 & Overt/Mimed & Speech Audio & 256 ch | 460 sentences | Open & \textbf{Two-stage bLSTM:} Decoded articulatory kinematics to synthesize high-intelligibility audio. 43\%--70\% word identification accuracy in listening tests. \\

\cite{metzger2023high} & ECoG (vSMC / STG / IFG) & 1 (Anarthria) & Attempted & Text / Audio / Avatar & 253 ch | Multi-sess | 1,024 words & \textbf{RNN + CTC + Vocoder:} 78 WPM median speed; 9.1\% WER (50-word) and 23.9\% WER (1,024-word). Integrated real-time speech synthesis and facial avatar. \\

\cite{chen2024neural} & ECoG (Bilateral STG) & 82 & Overt & Speech Waveform & 82 subj | 20.2h total | Open & \textbf{Neural2Speech (Transformer):} Large-scale self-supervised reconstruction. Mean Pearson $r = 0.81$; STOI = 0.52. Demonstrated performance gains from multi-patient scaling. \\

\cite{Littlejohn2025StreamingBrainToVoice} & ECoG (vSMC/STG/IFG) & 1 (Anarthria) & Silent Attempted & Text / Audio / Avatar & 253 ch | Multi-sess | 1,024 words & \textbf{RNN-T + Vocoder:} First continuously streaming speech BCI. 47.5--90.9 WPM; $<$1.0s latency. Utilized 80-ms windowing for near-instantaneous synthesis. \\

\midrule
\rowcolor{rowblue}
\multicolumn{7}{l}{
\rule[0.4ex]{0pt}{3.2ex}
\raisebox{0.3ex}{\textbf{Intracortical Microelectrode Arrays (MEA)}}
} \\[0.4ex]

\cite{Willett2021Intracortical} & MEA (BA 6v) & 1 & Attempted Handwriting & Characters & 192 ch | 5 sess | Alphabet+Symbols & \textbf{RNN (GRU) + LM:} Re-framed typing as a high-dimensional motor task. Achieved 90 CPM; 94.1\% raw character accuracy ($>$99\% with LM). \\

\cite{Willett2023SpeechBCI} & MEA (BA 6v) & 1 (ALS) & Attempted & Phonemes / Text & 128 ch | 10,850 sent. | 125k words & \textbf{RNN + CTC + LM:} High-performance text decoding. Achieved 62 WPM; 9.1\% WER (50-word) and 23.8\% WER (125,000-word). \\

\cite{card2024accurate} & MEA (vPCG) & 1 (ALS) & Attempted & Phonemes / Text & 256 ch | 84 sess (16h) | 125k words & \textbf{RNN + LM:} Rapidly calibrating neuroprosthesis. 32 WPM speed; 2.5\% WER (50-word) and 9.1\% WER (125,000-word). Performance stable over 8 months. \\

\cite{wairagkar2025instantaneous} & MEA (vPCG) & 1 (ALS) & Attempted & Acoustic Feat. & 256 ch | 4k--8k trials | Open & \textbf{Transformer + Vocoder:} First instantaneous causal voice synthesis. 10ms processing latency; 94.3\% mean transcript matching accuracy. \\

\cite{feghhi2025time} & MEA (BA 6v) & 1 (ALS) & Attempted & Phonemes / Text & 128 ch | B2T Benchmark | Open & \textbf{Time-Masked Transformer:} Optimized for real-time efficiency. 20.2\% relative WER reduction; 83\% fewer parameters and 52\% less GPU memory than GRU. \\

\cite{zhangcross} & MEA (Ventral Motor) & 2H + 7 NHP & Attempted / Imagined & Text & 128--256 ch | 367h total | Open & \textbf{Transformer + Audio-LLM:} Neural foundation model (BIT) pretrained on 367h of cross-species motor activity. Achieved SOTA end-to-end WER of 10.22\%. \\

\cite{Fogg2026generalizable} & MEA (vPCG) & 6 (ALS/stroke) & Attempted & Phonemes / Text & 1--6 arrays/user | Multi-user | $<$200 sent. adaptation & \textbf{Multi-user Transformer + LM:} Joint training across six participants reduced WER by 51.1\% on average relative to single-user models. Few-shot adaptation to a held-out user achieved $<$7\% WER with fewer than 200 sentences. \\

\cite{olak2026decoding} & MEA (BA 6v) & 1 (ALS) & Attempted & Phonemes / Text / MFCC & 128 ch | 24 sess | 12k sentences & \textbf{Transformer Seq2Seq + NHS:} SOTA sublexical readout; 14.3\% PER. Word decoding: 25.6\% WER (direct), 19.4\% WER (rescored). Demonstrated temporal chunking via attention maps. \\

\midrule
\rowcolor{rowblue}
\multicolumn{7}{l}{
\rule[0.4ex]{0pt}{3.2ex}
\raisebox{0.3ex}{\textbf{Non-Invasive and Multi-modal Frontiers}}
}\\[0.4ex]

\cite{tang2023semantic} & fMRI (Semantic Sys.) & 3 & Perc. / Imag. / Visual & Semantic & BOLD | 16h/subj | Stories & \textbf{GPT-1 Encoding Model:} First non-invasive reconstruction of continuous semantic narratives. Recovers the ``gist'' of perceived/imagined speech and silent videos. \\

\cite{defossez2023decoding} & MEG/EEG (Scalp) & 175 & Perceived & Speech Segments & 273 ch (MEG) | 163 h | 1,500+ Seg. & \textbf{Contrastive Learning:} Zero-shot decoding via wav2vec 2.0 alignment. Achieved 41\% avg. top-1 accuracy for MEG (80\% max). \\

\cite{zhang2024brain} & ECoG (vSMC) & 5 & Overt/ Whisper & Text & 256 ch | 160 sent. | 40 characters & \textbf{Modular CNN-RNN + LM:} First continuous Mandarin brain-to-text. 21\% avg. WER (14\% best); 93\% tone decoding accuracy. \\

\cite{Huth2016SemanticMaps} & fMRI (Whole) & 7 & Perception (Narrative) & Semantic Domains & Voxel-wise | $>$2h/subj | 985 feat. & \textbf{Voxel-wise Encoding:} Mapped intricate semantic ``atlases'' tiling the cortex. Identified 4 shared dimensions and 12 distinct category clusters consistent across individuals. \\

\bottomrule

\end{tabularx}

\vspace{0.2em}

{\fontsize{5.4pt}{5.2pt}\selectfont
\raggedright
\textit{Notes: $N$: number of subjects; \textbf{Anatomy:} vSMC = Ventral Sensorimotor Cortex; vPCG = Ventral Precentral Gyrus; STG = Superior Temporal Gyrus; IFG = Inferior Frontal Gyrus; BA = Brodmann Area. \textbf{Modality:} ECoG = Electrocorticography; MEA = Microelectrode Array. \textbf{Metrics:} WER = Word Error Rate; WPM/CPM = Words/Characters Per Minute; STOI = Short-Time Objective Intelligibility. \textbf{Methods:} LM = Language Model; LLM = Large Language Model; SSL = Self-Supervised Learning; RNN = Recurrent Neural Network; CTC = Connectionist Temporal Classification; B2T = Brain-to-Text Benchmark.}
}

\end{tcolorbox}

\end{table*}


\section{Closed-loop speech BCIs and co-adaptation}\label{cl_BCI_co_adapt}

Speech-related activity is hierarchical and dynamic (Section~\ref{neural_substrates}), recording interfaces drift (Section~\ref{modal_hardware}), and representational choices trade off decodability, robustness, and agency (Section~\ref{speech_rep}). These constraints become most consequential in real-time use, where the decoder enters the user's sensorimotor and cognitive loop. Users perceive outputs and errors, alter their strategy and neural activity, and may co-adapt with the model. This interaction can improve control, but can also destabilize performance, amplify bias, and obscure what the system is decoding.

\begin{figure}[!h]
    \centering
    \includegraphics[width=0.9\linewidth]{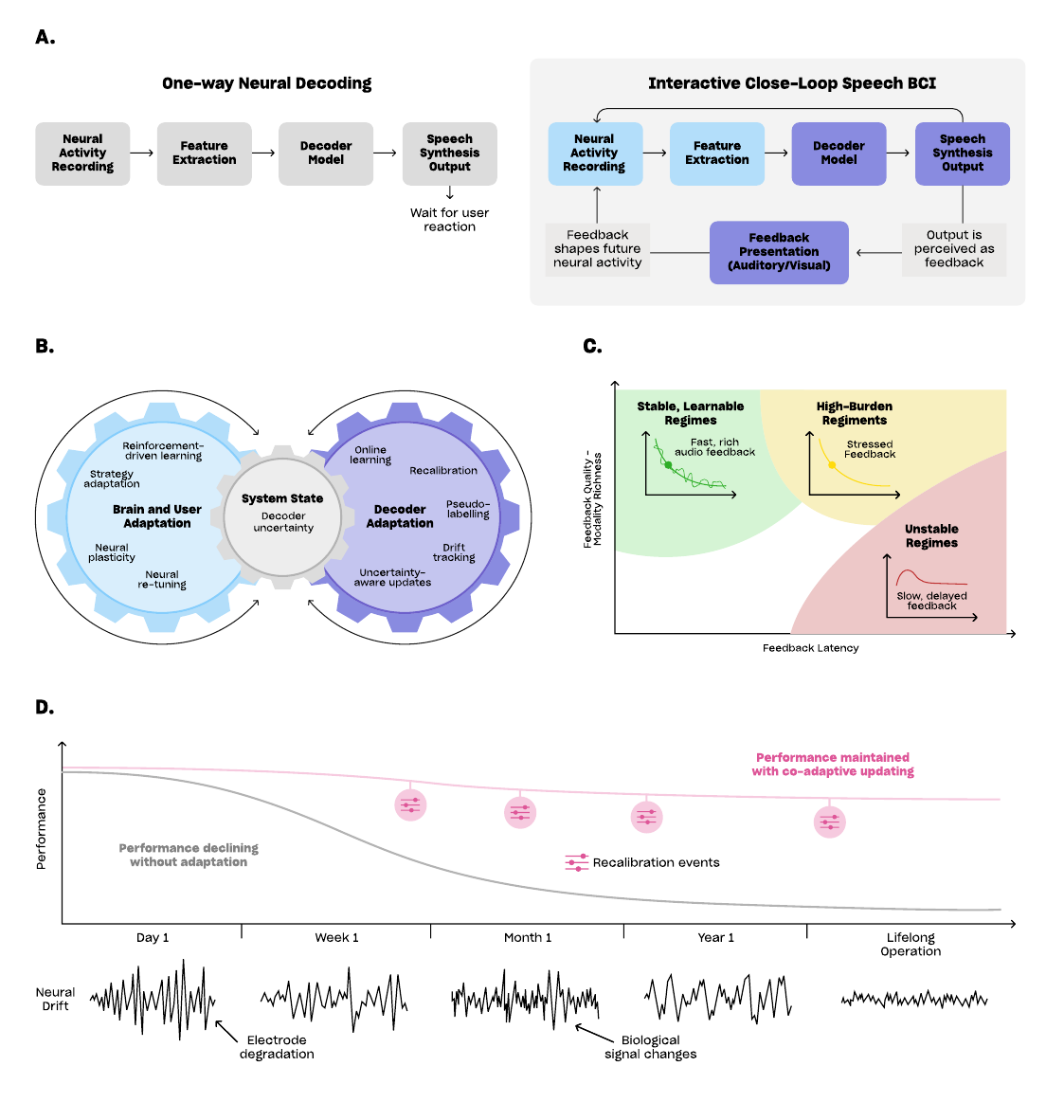}
    \caption{\captiontitle{Closed-loop co-adaptation in speech BCIs.}
    \textbf{A.} Conventional one-way decoding treats the decoder as a feedforward mapping from neural activity to speech output, with the user reacting only after the output is produced. In contrast, interactive closed-loop speech BCIs embed decoding within a feedback loop: decoded text or synthesized speech is perceived by the user and can shape subsequent neural activity, behavior, and control strategy.
    \textbf{B.} Co-adaptive operation arises from coupled brain/user adaptation and decoder adaptation. Brain/user adaptation includes neural plasticity, strategy adaptation, reinforcement-driven learning, and neural re-tuning, whereas decoder adaptation includes online learning, recalibration, drift tracking, uncertainty-aware updates, and pseudo-labeling/self-training. These processes interact through a shared system state defined by latency, feedback modality, decoder uncertainty, user burden, and stability.
    \textbf{C.} Feedback latency and feedback quality/modality richness shape distinct operating regimes. Low-latency, informative feedback supports stable and learnable interaction, shown as the green region. High-quality but delayed feedback can create high-burden regimes, shown in yellow, because rich feedback may still be difficult to use when temporal credit assignment is disrupted. Delayed and low-quality feedback can produce unstable regimes, shown in red, where control becomes unreliable and learning may degrade. The blank lower-left region represents fast but low-richness feedback: such feedback may be usable for simple correction or coarse control, but is not assumed to provide sufficient information for robust, stable speech learning.
    \textbf{D.} Across longer timescales, neural drift, electrode degradation, and biological signal changes can degrade performance when decoders remain fixed. Periodic recalibration and co-adaptive updating can help maintain performance during long-term or lifelong BCI operation.}
    \label{fig:4}
\end{figure}

Closed-loop learning is fundamental to BCI operation \cite{Wolpaw2002BCIOverview,Taylor2002LearningBCI,Ganguly2009Reorganization,orsborn2012closed}. Speech adds strong internal models, predictive mechanisms, and context-dependent feedback (Section~\ref{neural_substrates}); real-time inner-speech decoding further raises questions about intention, privacy, and control over when decoding occurs \cite{Kunz2025InnerSpeechCell}. Figure~\ref{fig:4} summarizes how brain adaptation, decoder adaptation, feedback, and drift jointly shape stability and long-term performance.


\subsection{User, decoder, and mutual adaptation}

\paragraph{Brain adaptation to the decoder.}
Users can learn to modulate neural activity even with a fixed decoder through reinforcement, error-based learning, and reorganization of population dynamics \cite{Taylor2002LearningBCI,Ganguly2009Reorganization,Sadtler2014NeuralConstraints}. This learning is constrained by the neural manifold accessible to the decoder \cite{Sadtler2014NeuralConstraints}. In speech BCIs, users may alter timing, exaggerate decodable features, or shift between attempted articulation and phonological imagery. Such strategies can improve control without reproducing canonical speech representations, particularly when feedback reinforces a narrow neural control channel \cite{orsborn2012closed}. Imagined-speech studies further show that continuous feedback can improve controllability and reshape spectral and spatial tuning \cite{bhadra2025learning}. Brain adaptation should therefore be measured and leveraged rather than treated as nuisance variability.

\paragraph{Decoder adaptation to the brain.}\label{cl_decoder_adaptation}
Adaptation also occurs in the opposite direction. Decoders must track changes caused by recording drift, fatigue, medication, motivation, and user strategy (Section~\ref{modal_hardware}); otherwise performance often declines across sessions \cite{Sussillo2016NeuralDrift,perge2014reliability}. Supervised recalibration uses periodically collected labelled data \cite{Gilja2012ReFIT}, whereas semi-supervised approaches update from high-confidence pseudo-labels \cite{Degenhart2020Stability}. State-space methods can instead model gradual changes in neural tuning explicitly \cite{Shenoy2013StateSpaceBCI}.

Speech introduces ambiguous natural-language labels, cascading errors from language priors, and the risk that rapid model updates destabilize a learned user policy. Conservative, uncertainty-aware adaptation (Section~\ref{decoding_uncertainty}) can mitigate these risks \cite{lee2025brain,Gal2016DropoutUncertainty,Guo2017Calibration}. Longitudinal independent use further demonstrates that low calibration burden, maintenance, and robust updating are practical requirements rather than optional refinements \cite{card2026long}.

\paragraph{Mutual co-adaptation.}\label{cl_mutual_adaptation}
In practice, brain and decoder usually adapt together. This can accelerate learning, but also create unstable feedback loops and make improvements difficult to attribute to neural control, decoder updates, or stronger priors \cite{orsborn2012closed}. Mutual adaptation can be viewed as a coupled dynamical system in which neural activity and model parameters evolve under reward, error, and prediction signals \cite{Friston2010PredictiveCodingSpeech}. Because speech production is itself predictive and feedback-driven (Section~\ref{neural_substrates}), a decoder changes the feedback environment and may reshape the processes it measures. Computational models of user--decoder interaction therefore argue that the best decoder is not necessarily the most accurate offline, but the one that supports stable, learnable, low-burden control \cite{madduri2024modeling}.

Evaluation should consequently report learning curves and within-session change, distinguish user-driven from model-driven improvement, and separate neural evidence from language-model contributions \cite{tang2023semantic,Kunz2025InnerSpeechCell}. These distinctions are central to agency and interpretability.

\subsection{Feedback design: sensory channels, latency, and user burden}\label{cl_BCI_co_adapt_feedback}

Feedback provides the link through which this co-adaptation occurs and may be auditory, textual, phoneme-level, confidence-based, or multimodal. Its latency and informativeness determine learning, correction, workload, and conversational usability. Delays of even a few hundred milliseconds can disrupt speech flow and sensorimotor prediction \cite{Houde1998SpeechPerturbation,Tourville2008AuditoryFeedback}. Streaming neuroprostheses reduce this gap and make interaction feel more like speaking than typing \cite{moses2021neuroprosthesis,Littlejohn2025StreamingBrainToVoice}, although causal decoding and short context windows can reduce accuracy.

Auditory feedback is naturalistic but may increase cognitive load when errors are frequent; text is easier to inspect and correct but less conversational. Imagined- and inner-speech systems also require explicit control over when decoding is active \cite{Kunz2025InnerSpeechCell, bhadra2025learning}.Calibration time, fatigue, attention, and the burden associated with errors should therefore be treated as primary outcomes \cite{card2024accurate,felton2012mental}, as discussed further in Section~\ref{eval_benchmark_rep}.

\subsection{Stability, drift correction, and lifelong operation}\label{cl_stability}

Over longer timescales, successful use must withstand electrode changes, neural plasticity, cognitive-state variation, and context shifts \cite{perge2014reliability,Sussillo2016NeuralDrift}. Speech strategy adds another source of drift because overt attempt and imagery can differ in signal strength and representational structure (Section~\ref{speech_rep_modes}) \cite{martin2018decoding,Kunz2025InnerSpeechCell}.

Correction can occur at several levels: adaptive filtering and normalization at the signal level; more invariant targets at the representation level (Section~\ref{speech_rep}) \cite{Anumanchipalli2019SpeechSynthesis,Huth2016SemanticMaps}; recalibration and uncertainty-aware updating at the model level \cite{Degenhart2020Stability,Gal2016DropoutUncertainty,Guo2017Calibration}; and feedback designs that prevent user and decoder from continually ``chasing'' one another \cite{orsborn2012closed,madduri2024modeling}. Lifelong systems additionally require safe update mechanisms, monitoring for unintended decoding, and user-controlled gating, especially when inner speech is accessible \cite{Kunz2025InnerSpeechCell}. These requirements connect technical stability to the clinical infrastructure considered in Section~\ref{clinical_trans}.

Closed-loop co-adaptation is thus a defining property of speech BCIs, shaping learning, stability, generalization, and user experience. It must be evaluated together with representational choice (Section~\ref{speech_rep}) and model design (Section~\ref{decoding}), rather than as an afterthought to offline accuracy.

\section{Evaluation, benchmarking, and reporting standards}\label{eval_benchmark_rep}

Progress in speech BCI decoding reflects advances in model capacity (Section~\ref{decoding}) and neural recording (Section~\ref{modal_hardware}), but offline accuracy captures only part of clinical utility. Closed-loop co-adaptation (Section~\ref{cl_BCI_co_adapt}), non-stationarity (Section~\ref{modal_hardware}), and representation mismatch (Section~\ref{speech_rep}) can all separate benchmark performance from lived use. Evaluation must therefore address fidelity, latency, throughput, robustness, generalization, uncertainty, and user burden. Table~\ref{tab:evaluation_translation}A summarizes the principal evaluation dimensions and their common pitfalls.

\begin{table*}[!ht]

\centering

\caption{\captiontitle{Evaluation and clinical translation considerations for speech BCIs}}
\label{tab:evaluation_translation}

\begin{tcolorbox}[
    colback=tether-gray,
    colframe=tether-gray,
    boxrule=0pt,
    arc=5pt,
    left=15pt,
    right=15pt,
    top=15pt,
    bottom=15pt,
    width=\textwidth
]

\scriptsize
\renewcommand{\arraystretch}{1.08}
\setlength{\tabcolsep}{5pt}


{\sharpfont\bfseries A | Recommended evaluation metrics}

\vspace{0.5em}

\begin{tabularx}{\linewidth}{
    >{\hsize=0.75\hsize\linewidth=\hsize\raggedright\arraybackslash}X
    >{\hsize=0.95\hsize\linewidth=\hsize\raggedright\arraybackslash}X
    >{\hsize=1.10\hsize\linewidth=\hsize\raggedright\arraybackslash}X
    >{\hsize=1.20\hsize\linewidth=\hsize\raggedright\arraybackslash}X
}

\toprule

\textbf{Metric category} &
\textbf{Representative measures} &
\textbf{Why it matters for deployment} &
\textbf{Common pitfalls} \\

\midrule

Accuracy &
WER, CER, phoneme accuracy &
Measures baseline decoding fidelity &
Can be inflated by strong language priors and may poorly reflect real-world usability \\

Latency &
End-to-end delay, streaming lag &
Determines turn-taking, conversational flow, and perceived responsiveness &
Often underreported or measured without accounting for the full system pipeline \\

Information rate &
Characters/min, words/min &
Captures effective communication throughput &
Can obscure the semantic severity and downstream cost of errors \\

Robustness &
Cross-session performance, sensitivity to channel dropout &
Reflects resilience to real-world non-stationarity &
Rarely evaluated beyond controlled laboratory conditions \\

Calibration and uncertainty &
ECE, confidence--accuracy alignment, abstention rate &
Supports safe interaction, error awareness, and appropriate user trust &
Often omitted despite its importance for clinical deployment \\

Usability and workload &
NASA-TLX, SUS, structured qualitative feedback &
Captures burden, learnability, and suitability for sustained daily use &
Often treated as anecdotal rather than as a primary outcome \\

\bottomrule

\end{tabularx}

\vspace{1.2em}


{\sharpfont\bfseries B | Clinical translation across target populations}

\vspace{0.5em}

\begin{tabularx}{\linewidth}{
    >{\hsize=0.80\hsize\linewidth=\hsize\raggedright\arraybackslash}X
    >{\hsize=1.00\hsize\linewidth=\hsize\raggedright\arraybackslash}X
    >{\hsize=0.95\hsize\linewidth=\hsize\raggedright\arraybackslash}X
    >{\hsize=1.25\hsize\linewidth=\hsize\raggedright\arraybackslash}X
}

\toprule

\textbf{Target population} &
\textbf{Clinically feasible modality} &
\textbf{Primary output form} &
\textbf{Dominant translational constraints} \\

\midrule

ALS (progressive) &
Intracortical arrays, ECoG &
Text or synthesized speech &
Rapid onboarding, minimal recalibration, fatigue management, disease progression \\

Brainstem stroke / locked-in syndrome &
Intracortical arrays &
Text or synthesized speech &
Surgical risk, chronic signal stability, long-term support \\

Spinal cord injury &
Intracortical arrays, ECoG &
Text or synthesized speech &
Sustained independent use, maintenance, software longevity \\

Cerebral palsy &
ECoG; EEG (limited feasibility) &
Text &
Anatomical heterogeneity, developmental reorganization, variable motor control \\

Non-invasive users &
EEG, MEG &
Text &
Low SNR, setup burden, caregiver dependence, limited throughput \\

\bottomrule

\end{tabularx}

\end{tcolorbox}

\end{table*}

\subsection{Metrics beyond accuracy: latency, information rate, and usability}\label{eval_metrics}

Phoneme or word accuracy, character error rate, word error rate, and sentence correctness remain essential measures of decoding fidelity \cite{herff2015brain,moses2021neuroprosthesis,Willett2023SpeechBCI}, but they should not stand alone. Conversational use depends on end-to-end latency because delays disrupt turn-taking, sensorimotor prediction, and learning (Section~\ref{cl_BCI_co_adapt_feedback}) \cite{Houde1998SpeechPerturbation,Tourville2008AuditoryFeedback}. Studies should therefore decompose delay into acquisition, preprocessing, inference, language-model integration, and rendering rather than report only per-window computation \cite{Littlejohn2025StreamingBrainToVoice}.

Throughput should likewise be reported as words or characters per minute, ideally after accounting for correction and error costs \cite{Wolpaw2002BCIOverview}. Aggregate rates can conceal severe semantic failures: a phonetic substitution may be recoverable, whereas a plausible but unintended semantic substitution can alter agency and safety \cite{tang2023semantic,sankaran2023recommendations}. Error analyses should therefore distinguish phonetic, lexical, and semantic consequences and quantify the contribution of downstream priors.

Usability and workload are primary clinical outcomes, not ancillary observations. Instruments such as the NASA Task Load Index and System Usability Scale, together with structured interviews, can assess fatigue, frustration, perceived control, and sustained acceptability \cite{Hart1988NASATLX,Brooke1996SUS}. Calibration time, attention demands, correction effort, and the emotional cost of mis-decoding should be reported alongside accuracy \cite{felton2012mental,card2024accurate}.

\subsection{Offline versus online performance}

Offline analyses can overestimate closed-loop performance because they often use segmented trials, stable task structure, future context, or non-causal processing \cite{orsborn2012closed}. Online operation adds strict latency constraints, variable cognitive state, and reciprocal adaptation between user and decoder \cite{Ganguly2009Reorganization,Sussillo2016NeuralDrift}. Clinically meaningful evaluation should therefore prioritize online communication, including learning curves, stability under drift, correction behaviour, and naturalistic interaction. When full closed-loop testing is infeasible, evaluation should approximate deployment through causal decoding, realistic buffering, and interactive correction \cite{moses2021neuroprosthesis,Willett2023SpeechBCI}.

Offline studies should also document segmentation, preprocessing statistics, validation procedures, and participant-specific language-model adaptation. Leakage can arise when test data influence normalization, alignment, hyperparameter selection, or prior tuning; confidence and calibration analyses do not repair an invalid split \cite{Guo2017Calibration}.

\subsection{Generalization across sessions, tasks, and subjects}

Generalization is a defining challenge because neural features and user strategies change across days \cite{perge2014reliability,Sussillo2016NeuralDrift}, while electrode coverage (Section~\ref{modal_hardware}) and representational mismatch (Section~\ref{speech_rep_mismatch}) vary across users. Longitudinal studies should report performance without recalibration, after minimal recalibration, and the time and data required for recovery. Where online adaptation is used (Section~\ref{cl_decoder_adaptation}), its update rules and stability safeguards should be specified.

Task generalization should include shifts from prompted to spontaneous speech, words to sentences, and overt to imagined production (Section~\ref{speech_rep_modes}), with explicit analysis of failure modes \cite{martin2018decoding,cooney2021bimodal}. Cross-subject transfer is harder because anatomy, pathology, language background, and electrode placement differ. Anatomical or functional normalization, shared latent spaces, and alignment to higher-level representations may reduce this mismatch \cite{Huth2016SemanticMaps,perge2013intra,sussillo2016making,singh2025transfer}. Studies should state whether training is subject-specific, pooled, or transfer-based and report participant-level dispersion rather than cohort averages alone.


\subsection{Robustness to noise, artifacts, and non-stationarity}

Robustness should cover measurement noise, channel loss, movement and muscle artifacts, impedance changes, fatigue, and cognitive-state variation. Speech tasks are especially vulnerable to facial, jaw, and auditory confounds, particularly in non-invasive recordings \cite{Lalor2009EEGSpeech}. Studies should document filtering, referencing, rejection criteria, and controls for non-neural signals, including EMG contamination and auditory-feedback timing \cite{Tourville2008AuditoryFeedback}.

Claims of robustness are stronger when supported by stress tests such as channel removal, additive noise, temporal jitter, and cross-session distribution shifts \cite{Degenhart2020Stability}. Calibration and uncertainty (Section~\ref{decoding_uncertainty}) should be evaluated under these perturbations so that confidence tracks correctness and enables abstention or fallback modes \cite{Gal2016DropoutUncertainty,Guo2017Calibration,Angelopoulos2023Conformal}.

\subsection{Datasets, benchmarks, and reproducibility checklists}\label{eval_data_bench}

Shared benchmarks remain limited by small cohorts, heterogeneous protocols, privacy constraints, and clinically determined electrode placement. Nevertheless, benchmark tasks should emphasize cross-session decoding, limited supervision, causal streaming, and realistic calibration rather than only within-session accuracy. Where data cannot be shared, structured dataset documentation can still expose participant characteristics, recording coverage, task design, annotations, preprocessing, and known limitations.

At minimum, reports should describe diagnosis, severity, inclusion criteria, medication, and relevant cognitive status \cite{FDAImplantedBCIGuidance2021}; modality, electrode geometry and coverage, sampling, referencing, and signal quality \cite{Leuthardt2004ECoG,perge2014reliability}; and whether speech was prompted, spontaneous, overt, mouthed, or imagined, including alignment and label uncertainty \cite{martin2018decoding}. Model reports should specify architecture, preprocessing, splits, hyperparameter selection, and leakage prevention. For systems using language priors, authors should disclose training data, participant adaptation, and ablations separating prior-driven from neurally supported performance \cite{tang2023semantic,Willett2023SpeechBCI}. Evaluation should additionally cover online and offline performance, latency, calibration, stress tests, workload, and usability \cite{Hart1988NASATLX,Brooke1996SUS}, together with code, data-access conditions, environments, and licensing where feasible. Box~\ref{box:reporting_ethics}A summarizes these elements.












\begin{reviewbox}[box:reporting_ethics]{Recommended reporting and responsible-design checklist for speech BCIs}

Transparent and responsible speech BCI research requires both comprehensive reporting of experimental and computational methods and explicit safeguards in system design.

\medskip
\noindent{\sharpfont\bfseries A | Reporting and reproducibility}

\medskip
\noindent\textbf{Participants and clinical context.}
Diagnosis, severity, inclusion criteria, medication status, and relevant cognitive assessments.

\medskip
\noindent\textbf{Recording details.}
Modality, electrode type and coverage, sampling rate, referencing scheme, and signal quality metrics.

\medskip
\noindent\textbf{Task design and labels.}
Prompted versus spontaneous speech, overt/mouthed/imagined conditions, alignment procedures, and label uncertainty.

\medskip
\noindent\textbf{Model and training.}
Architecture, preprocessing, data splits, hyperparameter selection, and leakage prevention.

\medskip
\noindent\textbf{Use of priors.}
Language model type, training data, participant adaptation, and ablations quantifying reliance on priors.

\medskip
\noindent\textbf{Evaluation.}
Offline and online metrics, latency, calibration, robustness tests, and user burden measures.

\medskip
\noindent\textbf{Availability.}
Code, data access conditions, and documentation of limitations.

\bigskip
\noindent{\sharpfont\bfseries B | Ethical and responsible design}

\medskip
\noindent\textbf{User-controlled activation.}
Explicit, reliable mechanisms that ensure decoding occurs only when the user intends to communicate.

\medskip
\noindent\textbf{Privacy-by-design.}
Minimization of continuous recording, secure storage, encryption, and access control for neural data, with local on-device decoding wherever feasible so that raw brain activity does not need to be transmitted to centralized servers.

\medskip
\noindent\textbf{Transparency of priors.}
Clear documentation of language-model use, provenance of output, and ablations quantifying reliance on priors.

\medskip
\noindent\textbf{Calibrated uncertainty.}
Confidence signalling, abstention mechanisms, and conservative defaults for high-impact outputs.

\medskip
\noindent\textbf{Bias auditing.}
Evaluation across languages, clinical conditions, and demographic factors, with explicit reporting of limitations.

\medskip
\noindent\textbf{Lifecycle accountability.}
Governance for software updates, model adaptation, and post-deployment monitoring.

\end{reviewbox}

\vspace{0.5em}
Evaluation practices determine what the field optimizes. Benchmarks centred on constrained offline accuracy can overstate progress, whereas metrics that incorporate latency, robustness, generalization, uncertainty, and user burden better reflect clinical viability. The next section considers how these requirements extend to naturalistic communication, including conversation, context, personalization, and multilingual use.

\section{Toward naturalistic communication}\label{nat_comms}

Most speech BCI demonstrations still use prompted vocabularies, short utterances, fixed timing, and structured laboratory tasks. Naturalistic communication instead requires fluid, expressive, context-sensitive interaction that supports real conversational goals with low burden and preserved agency. This depends jointly on interactive speech neuroscience (Section~\ref{neural_substrates}), long-term recording constraints (Section~\ref{modal_hardware}), representational choices (Section~\ref{speech_rep}), sequence models and priors (Section~\ref{decoding}), and closed-loop adaptation (Section~\ref{cl_BCI_co_adapt}). The studies reviewed in Section~\ref{empirical_landscape} show progress toward attempted speech, synthesis, and expressive output, but dialogue remains a central unmet goal. Moving from decoding to communication therefore requires treating the interface as an adaptive human--machine system rather than a one-way transcription tool.

\subsection{Conversational dynamics, context, and memory}

Natural conversation relies on cooperative turn-taking, with response gaps often lasting only a few hundred milliseconds despite substantial planning demands \cite{Levinson2016TurnTaking,LevinsonTorreira2015TimingTurnTaking,Meyer2023TimingConversation}. Speech BCIs must therefore operate incrementally, updating partial hypotheses rather than waiting for complete sentences. End-to-end latency should be treated as a primary outcome (Section~\ref{eval_metrics}), because even modest delays can disrupt turn transitions and conversational synchrony \cite{Levinson2016TurnTaking,Meyer2023TimingConversation}.

Incremental output also requires controlled commitment: uncertain words should remain revisable until sufficient evidence accumulates. Calibration, abstention, and confidence-aware display or speech (Section~\ref{decoding_uncertainty}) can reduce severe semantic errors without eliminating conversational flow \cite{Guo2017Calibration,Gal2016DropoutUncertainty}. Streaming neuroprostheses already produce words continuously rather than in delayed batches \cite{moses2021neuroprosthesis,Littlejohn2025StreamingBrainToVoice}, but natural dialogue additionally depends on hesitation, repair, backchannels, timing, and prosody. These interactive signals remain largely absent from current evaluations despite their importance for signalling attention, uncertainty, repair, and willingness to yield or retain the conversational floor.

Beyond timing, natural communication depends on dialogue history, shared knowledge, and pragmatic context. Language models can resolve ambiguity in low-level neural outputs, but may also generate fluent text weakly supported by neural evidence (Sections~\ref{decoding_ssl}--\ref{decoding_uncertainty}) \cite{moses2021neuroprosthesis,Willett2023SpeechBCI}. A useful architecture separates three components: a streaming neural likelihood over candidate units; a contextual language prior incorporating dialogue history, user-specific style, and only explicitly permitted external context; and a calibrated decision policy governing emission, revision, or abstention. This separation makes each contribution auditable and motivates prior ablations and counterfactual tests (Section~\ref{eval_data_bench}) \cite{Guo2017Calibration}.

Context also creates a memory-governance problem. Conversation history may be useful over minutes, personal vocabulary over days, and stable preferences over months, but persistence should be explicit, consented, auditable, and user-editable. Inner-speech decoding further requires user-controlled gating and clear transitions between active and inactive decoding states \cite{Kunz2025InnerSpeechCell}. Conversational memory should therefore not emerge opaquely from continual updates, and users should be able to inspect, correct, delete, or disable stored information.

\subsection{Personalization and multilingual communication}

\paragraph{Personalization and transfer.}
High-performing invasive BCIs are currently personalized because anatomy, electrode placement, and neural mappings vary substantially across individuals (Section~\ref{decoding_ssl}; Section~\ref{modal_hardware}) \cite{Sussillo2016NeuralDrift}. Personalization also supports co-adaptation between user and decoder (Section~\ref{cl_BCI_co_adapt}) \cite{orsborn2012closed,Ganguly2009Reorganization}, but imposes substantial calibration and maintenance costs.

Cross-user pretraining seeks to capture shared neural--behavioural structure and reduce this burden. Emerging intracortical results suggest that cross-brain transfer may be feasible with appropriate alignment and adaptation \cite{levin2026cross}, while non-invasive systems such as Brain2Qwerty illustrate how larger healthy-participant datasets may support scalable pretraining \cite{Levy2025Brain2Qwerty}. A plausible compromise is hybrid personalization: a general pretrained backbone with lightweight user-specific adapters, updated conservatively under uncertainty (Sections~\ref{decoding_ssl}--\ref{decoding_uncertainty}, \ref{cl_decoder_adaptation}) \cite{Degenhart2020Stability,Gal2016DropoutUncertainty}. Such systems should be judged by reduced calibration, robustness under longitudinal drift, and durable transfer, not merely by pretrained scale or average cross-user accuracy.

\paragraph{Multilingual communication.}\label{nat_multilingual}
Personalization must also accommodate linguistic diversity. Most speech BCI research has focused on English, although phoneme inventories, syllable structure, morphology, orthography, and prosody vary widely across languages (Section~\ref{speech_rep}). Tonal languages add lexical pitch and dense homophony, increasing demands on laryngeal control and contextual disambiguation \cite{liu2023decoding,zhang2024brain}. Intracranial studies have nevertheless decoded tonal speech and tone for brain-to-text and brain-to-speech pipelines \cite{liu2023decoding,zhang2024brain,qian2025real,qian2025real}.

Bilingual neuroprostheses indicate that shared articulatory representations can support more than one language while output mappings remain language-specific \cite{silva2024bilingual}. Multilingual decoding frameworks further suggest that shared semantic spaces may support cross-lingual transfer, including for low-resource languages \cite{guo2025pre}. Evaluation should therefore include code-switching, tone contrasts, morphologically diverse languages, and participant-level analysis across linguistic backgrounds. Language priors should remain configurable, with their training data, adaptation, and safeguards reported explicitly (Sections~\ref{eval_benchmark_rep}, \ref{clinical_trans}).

\subsection{Human--AI co-expression and interface design}\label{nat_coexpression}

Communication conveys prosody, emphasis, affect, timing, and identity as well as words. Expressive neural speech, including intonation and pitch control, demonstrates that these dimensions are technically accessible \cite{wairagkar2025instantaneous} and supports treating prosody and affect as primary targets (Section~\ref{speech_rep_prosody}).

Greater expressivity also blurs the boundary between decoding and generation. When an AI system selects prosody, paraphrases, or predicts continuations, it becomes a co-author. Interfaces should therefore provide rapid correction and rollback, distinguish neurally supported content from model inference, and offer granular control over activation and permitted outputs, particularly for inner speech \cite{Kunz2025InnerSpeechCell}. Personalization should preserve the user's vocabulary, tone, and self-presentation rather than impose a generic voice. Because users learn not only to control the decoder but also to communicate through this coupled channel (Section~\ref{cl_mutual_adaptation}) \cite{orsborn2012closed}, transparency and agency are integral to learning and trust, not merely ethical add-ons.

Naturalistic speech BCIs must therefore optimize interactive dialogue rather than offline transcription alone: low-latency incremental decoding, calibrated contextual fusion, efficient personalization, multilingual robustness, expressive output, and explicit agency safeguards. The next section considers the ethical, legal, and societal implications of increasingly naturalistic and continuously available communication systems.

\section{Clinical translation and deployment}\label{clinical_trans}

The scientific questions that animate speech BCI research---what can be decoded, from where, and with what models (Sections~\ref{neural_substrates}--\ref{decoding})---ultimately converge on whether these systems can restore communication safely, reliably, and equitably in daily life. Translation requires clinically meaningful endpoints, low-burden training and support, maintainability under drift (Section~\ref{cl_BCI_co_adapt}), and regulatory oversight of both implantable hardware and its software stack \cite{Wolpaw2002BCIOverview,FDAImplantedBCIGuidance2021,IMDRF_SaMD_N41_2017}.

Deployment is shaped by heterogeneous clinical presentations, caregiver involvement, long-term maintenance, data governance, cybersecurity, and failure handling---constraints often underrepresented in academic evaluation (Section~\ref{eval_benchmark_rep}). Recent communication neuroprostheses show that practical success depends as much on system integration and operational reliability as on decoder accuracy \cite{moses2021neuroprosthesis,Willett2023SpeechBCI}. Across implantable BCIs, inconsistent and weakly standardized clinical outcomes remain a major barrier to comparison and regulatory translation \cite{dohle2025toward}. Table~\ref{tab:evaluation_translation}B summarizes representative target populations, feasible recording modalities, communication outputs, and deployment constraints.




















\subsection{Target populations and clinical endpoints}

Speech BCIs primarily target people with preserved communicative intent but severely impaired motor output, including ALS, brainstem stroke, spinal cord injury, advanced cerebral palsy, and near locked-in states \cite{card2024accurate}. Disease progression, residual movement, cognition, and home support differ substantially across these groups and determine feasible interfaces and training schedules.

Endpoints should therefore measure functional communication rather than idealized offline accuracy alone. Beyond WER or CER (Section~\ref{eval_benchmark_rep}), studies should quantify reliability across days, correction and breakdown recovery, and real-world communication efficiency \cite{Wolpaw2018HomeUseNeurology}. Autonomy should include independent setup, calibration, and use, together with reduced caregiver dependence \cite{card2026long}; patient-reported communication participation and quality of life are also important. Safety assessment must cover perioperative risk and long-term device complications \cite{FDAImplantedBCIGuidance2021}. Priorities should match the intended use: progressive disease may demand rapid onboarding and minimal recalibration, whereas stable injury may permit longer training; synthesized speech emphasizes latency and intelligibility, while brain-to-text additionally requires efficient correction, privacy controls, and sustained home use \cite{moses2021neuroprosthesis,Willett2023SpeechBCI}. Endpoint heterogeneity across implantable BCI studies reinforces the need for population- and function-specific standards \cite{dohle2025toward}.

\subsection{Deployment burden, reliability, and accessibility}

Training and calibration consume limited clinical resources and compete with fatigue, attention, and caregiver availability. Even accurate systems may be impractical if they require frequent, lengthy recalibration \cite{card2024accurate}; home systems must also accommodate daily changes in medication, sleep, and cognitive state (Section~\ref{cl_BCI_co_adapt}).

Caregiver burden also differs by modality. Non-invasive systems may require cap placement, electrode preparation, and troubleshooting \cite{Wolpaw2018HomeUseNeurology}, whereas implants shift burden toward charging, device management, software updates, and technical support. Accessibility further depends on surgical eligibility, specialist centres, cost and reimbursement, literacy, language support, and assistive-technology training. English-centric language models can compound these inequities (Section~\ref{nat_multilingual}). Multilingual interfaces, adaptable controls, and caregiver-inclusive training should therefore be designed from the outset.

Beyond initial training and access, sustained deployment requires reliable operation under changing signals and real-world conditions. Because long-term drift is expected (Section~\ref{cl_stability}), the relevant criterion is stable communication under realistic maintenance schedules rather than invariant signals \cite{perge2014reliability,Sussillo2016NeuralDrift}. Sustained use also requires a service model for consented remote monitoring, clinical follow-up, safe updates, and troubleshooting. Long-term independent use demonstrates feasibility, but also the importance of engineering and support beyond the offline decoder \cite{card2026long,Wolpaw2018HomeUseNeurology}.

Systems should also fail predictably: abstain when uncertain (Section~\ref{decoding_uncertainty}), offer low-bandwidth fallback modes, and provide user-controlled activation, particularly as inner-speech decoding improves \cite{Gal2016DropoutUncertainty,Guo2017Calibration,Kunz2025InnerSpeechCell}. Deployment favours streaming-capable, channel-robust, monitorable models that can be updated within regulated software lifecycles \cite{IMDRF_SaMD_N41_2017}, rather than computationally intensive models dependent on non-causal context or sensitive external data.

\subsection{Regulatory pathways and safety considerations}

In the United States, implantable BCIs are regulated as medical devices, and early human studies typically proceed under an Investigational Device Exemption, including early feasibility pathways for significant-risk devices \cite{FDAEarlyFeasibilityIDE2018,FDAImplantedBCIGuidance2021}. Dedicated FDA guidance addresses preclinical testing, biocompatibility, electromagnetic compatibility, packaging, human factors, and risk analysis \cite{FDAImplantedBCIGuidance2021}; recent speech-restoration IDE clearances indicate accelerating translation \cite{Paradromics2025ConnectOneIDE,Webb2025MedtechInsightParadromicsIDE}.

Safety extends beyond implantation. Output safety requires preventing unintended or misattributed communication, enabling user-controlled activation, and displaying confidence and revision states (Sections~\ref{decoding_uncertainty} and \ref{nat_coexpression}) \cite{Kunz2025InnerSpeechCell,Guo2017Calibration}. Human-factors engineering must reduce workload and error cascades \cite{FDAImplantedBCIGuidance2021}, while cybersecurity and data protection are essential for connected systems handling sensitive neural data \cite{FDA_Cybersecurity_Guidance_2025}.

Because many risks arise from software updates, translation also requires lifecycle governance. ISO~14971 and IMDRF software-as-a-medical-device frameworks provide structures for hazard identification, clinical evaluation, risk control, and post-deployment monitoring \cite{ISO14971_2019,IMDRF_SaMD_N41_2017,IMDRF_SaMD_N23_2015}. A central unresolved issue is how continual adaptation (Section~\ref{cl_BCI_co_adapt}) can coexist with regulated change control. Near-term systems may therefore favour conservative updates, locked versions, and explicit recalibration protocols over autonomous continual learning.

Clinical translation requires rebalancing priorities from peak accuracy toward safe, reliable, equitable, and low-burden operation under realistic constraints. The next section addresses the ethical, legal, and societal implications of cognitive privacy, consent, agency, and access as speech BCIs move into everyday communication.

\section{Ethical, legal, and societal implications}\label{ethic_legal}

As speech BCIs move toward clinical deployment (Section~\ref{clinical_trans}) and naturalistic communication (Section~\ref{nat_comms}), ethical, legal, and societal questions become part of system design. These interfaces may access internally generated language and increasingly combine neural evidence with adaptive language models, complicating agency, authorship, accountability, and bias (Sections~\ref{decoding}--\ref{eval_benchmark_rep}). Emerging governance frameworks therefore emphasize mental privacy, freedom of thought, and lifecycle responsibility \cite{OECD2019NeurotechRecommendation,UNESCO2025NeurotechEthics,Ruiz2024Neurorights,Spichak2025NeurorightsJMIR}. The practical challenge is to translate these principles into user control, transparent model behaviour, equitable development, and accountable deployment. Box~\ref{box:reporting_ethics}B summarizes core design requirements.


\subsection{Privacy of neural data and inner speech}

Neural recordings may reveal information beyond intended communication, including attention, affect, and internally generated language. This sensitivity will increase with higher-channel-count implants, continuous monitoring, and stronger models (Sections~\ref{modal_hardware} and~\ref{decoding}). Neural data should therefore be treated as highly sensitive throughout collection, storage, processing, and sharing \cite{OECD2019NeurotechRecommendation,UNESCO2025NeurotechEthics,OECDNeurotechToolkit2024}.

Empirical decoding of covert or inner speech sharpens concerns about mental privacy and the boundary between thought and communication \cite{Kunz2025InnerSpeechCell,almufareh2025inner}. Outputs must remain \emph{user-authorized}: decoding should occur only when the user intends to communicate, with reliable activation and deactivation \cite{Kunz2025InnerSpeechCell}. Layered controls may combine a voluntary ``push-to-talk'' signal with confirmation for low-confidence or high-impact outputs (Section~\ref{decoding_uncertainty}). Privacy-by-design should additionally minimize collection and retention, secure storage and processing, and restrict access \cite{OECD2019NeurotechRecommendation,UNESCO2025NeurotechEthics}.

Consent must also persist across the system lifecycle. Users should be able to review and revise permissions as models, interfaces, and downstream uses evolve, including separate choices for attempted versus covert speech, text versus audio output, conversational memory, and cloud processing \cite{UNESCO2025NeurotechEthics}. Researchers should clearly state what is and is not being decoded to avoid both exaggerating capabilities and understating risks \cite{almufareh2025inner}.

\subsection{Agency, authorship, and responsibility}

When speech BCIs fuse neural evidence with language-model priors (Sections~\ref{decoding} and~\ref{nat_comms}), fluent output may be partly inferred rather than directly decoded. Interfaces should preserve authorship by exposing confidence and revision states, enabling rapid rollback, and allowing users to limit model-driven completion (Section~\ref{nat_coexpression}). Systems that paraphrase, summarize, or predict continuations should be described as \emph{co-expression}, not literal transcription, with clear provenance and opt-out controls.

Accountability becomes difficult when erroneous output causes reputational, relational, or legal harm. Although responsibility may be distributed across users, clinicians, developers, and institutions, systems can reduce risk through calibrated uncertainty, abstention, and conservative defaults (Section~\ref{decoding_uncertainty}) \cite{OECD2019NeurotechRecommendation,UNESCO2025NeurotechEthics}. Studies should characterize phonetic and semantic errors, distinguish benign from high-impact failures, quantify whether language models amplify or conceal them, and test confirmation and veto mechanisms (Section~\ref{eval_metrics}) \cite{Guo2017Calibration,Willett2023SpeechBCI,tang2023semantic}.










\subsection{Bias, fairness, and representativeness}

Speech BCI bias can arise from small, clinically selected datasets; English-dominant language models; heterogeneous disease and device characteristics; and group-dependent calibration or confidence errors \cite{Guo2017Calibration}. Because participation depends on implant eligibility, specialist access, and tolerance for long studies, datasets may not represent the populations that could benefit. Studies should therefore report clinical and demographic context, stratify performance by relevant factors, and audit differential errors (Sections~\ref{eval_benchmark_rep} and~\ref{nat_comms}), including across languages and dialects (Section~\ref{nat_multilingual}). Clinical heterogeneity—including lesion location, disease progression, medication, fatigue, and cognitive state—can further produce systematic performance differences (Section~\ref{clinical_trans}). Evaluation should also examine whether thresholds, confidence estimates, and correction mechanisms work comparably across groups (Section~\ref{eval_benchmark_rep}). When data cannot be shared, datasheets and model cards can still document assumptions, exclusions, and limitations \cite{Gebru2018Datasheets,Mitchell2019ModelCards}.

Fairness also concerns access. Cost, surgical availability, caregiver training, multilingual support, and concentration in specialist centres may widen existing health disparities \cite{UNESCO2025NeurotechEthics,OECD2019NeurotechRecommendation}. Benchmark design should therefore include heterogeneous clinical cohorts and multilingual tasks rather than optimizing only on highly curated English datasets (Section~\ref{eval_benchmark_rep}).

\subsection{Neuro-rights and governance}

The neurorights movement argues that existing human-rights frameworks may not fully address threats to mental privacy, identity, and autonomy \cite{Ruiz2024Neurorights,Spichak2025NeurorightsJMIR}. Chile has become a prominent case through constitutional, legislative, and judicial developments concerning neurorights and mental privacy \cite{guzman2022chile,cornejo2023chilean}. Internationally, the OECD Recommendation on Responsible Innovation in Neurotechnology emphasizes stewardship, safety, privacy, trust, and inclusive deliberation \cite{OECD2019NeurotechRecommendation}, while UNESCO's 2025 Recommendation foregrounds dignity, freedom of thought, privacy, accountability, and lifecycle safeguards \cite{UNESCO2025NeurotechEthics}. Together, these initiatives imply stronger expectations for consent, data governance, and constraints on sensitive or non-therapeutic uses \cite{OECDNeurotechToolkit2024,UNESCO2025NeurotechEthics}.

For speech BCIs, governance should translate into user-controlled activation, minimized continuous recording, auditable logs, secure processing, and restricted retention \cite{OECD2019NeurotechRecommendation,UNESCO2025NeurotechEthics}. Developers should disclose language-model training and adaptation, and quantify how much output derives from priors rather than neural evidence \cite{Willett2023SpeechBCI,tang2023semantic,moses2021neuroprosthesis}. Lifecycle accountability should additionally govern software updates, model adaptation, and post-deployment monitoring as regulated medical software (Section~\ref{clinical_trans}) \cite{IMDRF_SaMD_N41_2017}.

Ethical and societal considerations are therefore not peripheral constraints: they determine what systems should decode, how outputs should be represented, and which deployment pathways are acceptable. The concluding section synthesizes the field's scientific and translational bottlenecks and identifies priorities for progress that preserve agency, privacy, and equitable benefit.


\section{Open challenges and future directions}

Speech BCIs aim to restore communication by transforming neural activity related to speech, language, or communicative intent into text, synthesized speech, or another controllable output. Their progress must be judged both by clinical utility and by the scientific insight they provide into speech and language. Yet the properties that make these systems powerful also make them difficult to deploy: invasive interfaces offer high-bandwidth signals but require surgery and long-term maintenance, whereas non-invasive systems are more accessible but remain limited in specificity, bandwidth, and real-time reliability. The major open challenges therefore concern which neural representations to decode, how to maintain robust and flexible models, and how to build sufficiently large and comparable datasets without compromising agency or privacy.

\begin{itemize}

\item \textbf{Scientific bottlenecks.}
There is no single neural representation of communication to decode. Speech-related activity varies across time, cortical space, behavioural context, and individuals. At short timescales, neural signals mix planning, articulatory execution, sensory prediction, feedback, and error monitoring; over longer periods, they change with electrode properties, plasticity, fatigue, cognitive state, and user strategy \cite{Sussillo2016NeuralDrift,perge2014reliability}. Different regions also emphasize different levels of the hierarchy: ventral sensorimotor and premotor areas often carry articulatory and motor-intent information, superior temporal regions represent acoustic and phonetic structure, and distributed temporal, parietal, and frontal networks support lexical, semantic, and contextual representations \cite{Bouchard2013MotorCortexSpeech,Mesgarani2014PhoneticFeatures,Huth2016SemanticMaps}. These mappings vary further with anatomy, pathology, electrode placement, language, and compensatory strategies.

A central question is therefore which representation, or combination of representations, best supports communication. Low-level articulatory and phonetic targets can enable fast, constrained decoding and have supported high-performance attempted-speech neuroprostheses \cite{Willett2023SpeechBCI,card2024accurate}. Higher-level semantic representations may be more distributed and potentially more stable, but are further removed from moment-to-moment intent and more susceptible to interpretation by language-model priors \cite{tang2023semantic}. Future systems may need hierarchical models that combine low-level evidence with higher-level context while preserving traceability between neural signals and emitted content. Accordingly, future implantation strategies should be representation-driven, using functional mapping to identify the distributed—and where relevant bilateral—neural populations best matched to the intended communication function.

Communication is also multimodal. People convey intent through speech, handwriting, typing, facial expression, prosody, gesture, gaze, and contextual cues. BCIs should therefore move beyond text transcription toward flexible communication repertoires that include synthesized voice, expressive prosody, avatar control, spelling, handwriting-like sequences, and other assistive outputs \cite{Willett2021Intracortical,metzger2023high,wairagkar2025instantaneous}. The ultimate target is communication rather than literal speech reconstruction: the optimal interface may exploit whichever neural representation or expressive channel provides the most reliable, controllable, and personally meaningful route from communicative intent to interaction.

Intent is the final defining scientific challenge. As decoders become sensitive to inner or imagined speech, they must distinguish activity intended for communication from internal monologue. Possible approaches include deliberate activation signals, intent detection integrated into the decoder, and confirmation before releasing uncertain or high-impact outputs \cite{Kunz2025InnerSpeechCell}. These mechanisms must be robust to false activation and remain easy to use under fatigue or limited motor control. Understanding how communicative intent is represented and changes during closed-loop use is therefore essential to building systems that are both powerful and trustworthy.

\item \textbf{Technical bottlenecks.}
The foremost technical challenge is maintaining performance under neural non-stationarity. Even with fixed hardware, mappings between neural features and intended outputs change across hours, days, and months because of drift, fatigue, medication, plasticity, impedance changes, and evolving strategies \cite{Sussillo2016NeuralDrift,perge2014reliability}. Future systems will require adaptive normalization, robustness to channel loss, uncertainty-aware updates, and online learning that tracks change without catastrophic forgetting or destabilizing the user's learned control policy.

Rapid self-calibration is closely related. Participant-specific retraining can become burdensome for users and caregivers, so practical systems should minimize daily labelled data through supervised micro-calibration, semi-supervised learning from high-confidence predictions, drift-aware state-space models, or lightweight user-specific adapters on pretrained encoders \cite{Gilja2012ReFIT,Degenhart2020Stability,Shenoy2013StateSpaceBCI}. Adaptation must remain conservative, because erroneous pseudo-labels can reinforce mistakes and cause the brain and decoder to chase one another.

Generalization across tasks is another unmet requirement. Current systems are usually optimized for prompted speech, phoneme classification, sentence reading, imagined speech, or another narrow paradigm. Everyday use requires transitions across spontaneous and prompted communication, vocabularies, output formats, languages, and cognitive strategies. Multi-task learning, task-conditioned decoders, shared latent spaces, and modular output heads may help, but must be evaluated under explicit task shifts rather than inferred from within-task accuracy.

Finally, language models must support rather than override neural evidence. They improve fluency and large-vocabulary decoding but can mask weak signals and bias output toward statistically common phrases \cite{tang2023semantic,Willett2023SpeechBCI}. This becomes especially problematic for unusual names, passwords, jokes, poetry, code-switching, or other low-probability content. Their influence should therefore be measurable, exposed, and adjustable through ablations, calibrated confidence, alternative hypotheses, and user-controlled correction. Technical progress should be assessed by sustained performance under drift, low calibration burden, cross-task flexibility, and transparent separation of neural evidence from linguistic priors rather than peak offline accuracy alone.

\item \textbf{Data and benchmarking bottlenecks.}
Speech BCI research remains constrained by data scarcity. Invasive datasets depend on rare clinical opportunities, individualized electrode placement, participant fatigue, and limited recording time, while high-capacity models may require thousands of labelled trials from a single participant. This mismatch encourages overfitting to user identity, task structure, or recording conditions and limits transfer across sessions, participants, and communication contexts.

Self-supervised and weakly supervised learning may reduce reliance on dense labels. Masked autoencoding, contrastive learning, predictive objectives, and joint embedding predictive architectures could learn reusable neural representations from unlabelled or weakly aligned recordings before task-specific fine-tuning \cite{defossez2023decoding,jayalath2024brain}. However, neural pretraining objectives can exploit shortcuts based on timing, stimulus identity, session structure, or artifacts. Positive and negative pairs, masking policies, context windows, and augmentations must therefore be designed around neural and behavioural structure and validated through stringent controls.

Combining heterogeneous datasets presents an additional problem. Participants differ in anatomy, pathology, language, and implant location; channels do not correspond naturally across individuals; and tasks may involve overt, attempted, imagined, or perceived speech. Naive pooling may erase meaningful distinctions rather than produce general representations. Promising directions include anatomical or functional alignment, shared latent spaces, subject- and site-specific adapters, task conditioning, models that tolerate missing channels, and privacy-preserving multi-site learning. Foundation models will be useful only if they demonstrably reduce calibration and improve transfer under these sources of heterogeneity, rather than merely increasing model scale.

Benchmarking must likewise move beyond heterogeneous within-session scores. Studies differ in vocabulary, recording duration, preprocessing, language-model use, latency, and whether evaluation is offline or online. Common benchmarks should test cross-session stability, low-calibration adaptation, channel loss, task transfer, causal streaming, closed-loop use, and separation of neural evidence from language priors. Participant-level results, calibration burden, and failure modes should accompany aggregate accuracy.

A long-term priority is to make progress cumulative rather than study-specific. Where raw neural data cannot be openly shared, standardized reporting, common protocols, model cards, dataset datasheets, and federated or multi-site frameworks can still improve comparison and reuse \cite{Gebru2018Datasheets,Mitchell2019ModelCards}. Larger datasets alone will not solve speech BCI decoding; they must be paired with inductive biases that reflect neural dynamics, speech hierarchy, communicative intent, and closed-loop interaction.

\end{itemize}

\section{Conclusion}

Speech BCIs have progressed from proof-of-concept demonstrations toward increasingly high-performance systems capable of decoding attempted speech, text, and synthesized voice. These advances show that neural activity can support meaningful restoration of communication for people with severe speech and motor impairments. Yet they also reveal that speech BCIs are not simply a problem of mapping neural signals to words. They are complex, adaptive systems in which neural representations, recording hardware, decoding architectures, language priors, feedback, and user learning interact over time.

A central message of this Review is that the future of speech BCIs depends on aligning decoding targets with the neural representations that are actually accessible, stable, and useful for communication. Articulatory, phonetic, acoustic, lexical, semantic, and pragmatic representations each offer different advantages and limitations. Similarly, invasive and non-invasive modalities occupy different positions in the trade-off between signal quality, clinical burden, spatial coverage, temporal resolution, and long-term feasibility. No single representation, model, or device will solve the problem in isolation; progress will require system-level designs that explicitly account for hierarchy, distribution, temporal dynamics, non-stationarity, and closed-loop co-adaptation.

The next generation of speech BCIs will therefore need to move beyond peak offline accuracy as the primary marker of progress. Clinically meaningful systems must be robust across sessions, rapidly calibratable, uncertainty-aware, low latency, and usable under the cognitive and practical constraints of daily life. They must also make the contribution of language models transparent, preserve the user's ability to communicate atypical or low-probability messages, and provide safeguards against unintended decoding. As speech BCIs approach inner speech, naturalistic dialogue, expressive synthesis, and multimodal communication, questions of agency, authorship, privacy, and equity become central design requirements rather than peripheral ethical concerns.

Speech BCIs now sit at a pivotal point: the field has demonstrated that restoring communication from brain activity is possible, but translating this possibility into reliable, scalable, and user-controlled communication remains an open scientific, technical, and clinical challenge. Meeting this challenge will require closer integration between speech neuroscience, neural engineering, machine learning, clinical practice, and neuroethics. If these strands are developed together, speech BCIs may not only restore a lost channel of communication, but also reshape our understanding of how intentions, language, and interaction are represented in the human brain.


\begin{thebibliography}{100}

\bibitem{Birbaumer2006BCIClinical}
N.~Birbaumer, ``Breaking the silence: brain--computer interfaces (bci) for communication and motor control,'' {\em Psychophysiology}, vol.~43, no.~6, pp.~517--532, 2006.

\bibitem{card2024accurate}
N.~S. Card, M.~Wairagkar, C.~Iacobacci, X.~Hou, T.~Singer-Clark, F.~R. Willett, E.~M. Kunz, C.~Fan, M.~Vahdati~Nia, D.~R. Deo, {\em et~al.}, ``An accurate and rapidly calibrating speech neuroprosthesis,'' {\em New England Journal of Medicine}, vol.~391, no.~7, pp.~609--618, 2024.

\bibitem{Wolpaw2002BCIOverview}
J.~R. Wolpaw, N.~Birbaumer, D.~J. McFarland, G.~Pfurtscheller, and T.~M. Vaughan, ``Brain–computer interfaces for communication and control,'' {\em Clinical Neurophysiology}, vol.~113, pp.~767--791, 2002.

\bibitem{Farwell1988P300}
L.~A. Farwell and E.~Donchin, ``Talking off the top of your head: Toward a mental prosthesis utilizing event-related brain potentials,'' {\em Electroencephalography and Clinical Neurophysiology}, vol.~70, pp.~510--523, 1988.

\bibitem{herff2015brain}
C.~Herff, D.~Heger, A.~De~Pesters, D.~Telaar, P.~Brunner, G.~Schalk, and T.~Schultz, ``Brain-to-text: decoding spoken phrases from phone representations in the brain,'' {\em Frontiers in neuroscience}, vol.~9, p.~217, 2015.

\bibitem{brumberg2010brain}
J.~S. Brumberg, A.~Nieto-Castanon, P.~R. Kennedy, and F.~H. Guenther, ``Brain--computer interfaces for speech communication,'' {\em Speech communication}, vol.~52, no.~4, pp.~367--379, 2010.

\bibitem{moses2021neuroprosthesis}
D.~A. Moses, S.~L. Metzger, J.~R. Liu, G.~K. Anumanchipalli, J.~G. Makin, P.~F. Sun, J.~Chartier, M.~E. Dougherty, P.~M. Liu, G.~M. Abrams, {\em et~al.}, ``Neuroprosthesis for decoding speech in a paralyzed person with anarthria,'' {\em New England Journal of Medicine}, vol.~385, no.~3, pp.~217--227, 2021.

\bibitem{Willett2023SpeechBCI}
F.~R. Willett, E.~M. Kunz, C.~Fan, D.~T. Avansino, G.~H. Wilson, E.~Y. Choi, F.~Kamdar, M.~F. Glasser, L.~R. Hochberg, J.~M. Henderson, and K.~V. Shenoy, ``A high-performance speech neuroprosthesis,'' {\em Nature}, vol.~620, pp.~103--111, 2023.

\bibitem{Sussillo2016NeuralDrift}
D.~Sussillo, M.~M. Churchland, M.~T. Kaufman, and K.~V. Shenoy, ``A neural network that finds a naturalistic solution for the production of muscle activity,'' {\em Nature Neuroscience}, vol.~18, pp.~1025--1033, 2015.

\bibitem{perge2014reliability}
J.~A. Perge, S.~Zhang, W.~Q. Malik, M.~L. Homer, S.~Cash, G.~Friehs, E.~N. Eskandar, J.~P. Donoghue, and L.~R. Hochberg, ``Reliability of directional information in unsorted spikes and local field potentials recorded in human motor cortex,'' {\em Journal of neural engineering}, vol.~11, no.~4, p.~046007, 2014.

\bibitem{Hickok2012SpeechNeuralBasis}
G.~Hickok and D.~Poeppel, ``The cortical organization of speech processing,'' {\em Nature Reviews Neuroscience}, vol.~8, pp.~393--402, 2007.

\bibitem{Price2012LanguageBrain}
C.~J. Price, ``A review and synthesis of the first 20 years of pet and fmri studies of heard speech, spoken language and reading,'' {\em NeuroImage}, vol.~62, pp.~816--847, 2012.

\bibitem{Pasley2012Reconstruction}
B.~N. Pasley, S.~V. David, N.~Mesgarani, A.~Flinker, S.~A. Shamma, N.~E. Crone, R.~T. Knight, and E.~F. Chang, ``Reconstructing speech from human auditory cortex,'' {\em PLoS Biology}, vol.~10, p.~e1001251, 2012.

\bibitem{Akbari2019SpeechReconstruction}
H.~Akbari, B.~Khalighinejad, J.~L. Herrero, A.~D. Mehta, and N.~Mesgarani, ``Towards reconstructing intelligible speech from the human auditory cortex,'' {\em Scientific Reports}, vol.~9, p.~874, 2019.

\bibitem{Moses2019SentenceDecoding}
D.~A. Moses, M.~K. Leonard, J.~G. Makin, and E.~F. Chang, ``Real-time decoding of question-and-answer speech dialogue using human cortical activity,'' {\em Nature communications}, vol.~10, no.~1, p.~3096, 2019.

\bibitem{zhang2025decoding}
Y.~Zhang, L.~He, C.~Fan, T.~Liu, H.~Yu, T.~Le, J.~Li, S.~Linderman, L.~Duncker, F.~R. Willett, {\em et~al.}, ``Decoding inner speech with an end-to-end brain-to-text neural interface,'' {\em arXiv preprint arXiv:2511.21740}, 2025.

\bibitem{tang2023semantic}
J.~Tang, A.~LeBel, S.~Jain, and A.~G. Huth, ``Semantic reconstruction of continuous language from non-invasive brain recordings,'' {\em Nature Neuroscience}, vol.~26, no.~5, pp.~858--866, 2023.

\bibitem{martin2018decoding}
S.~Martin, I.~Iturrate, J.~d.~R. Mill{\'a}n, R.~T. Knight, and B.~N. Pasley, ``Decoding inner speech using electrocorticography: Progress and challenges toward a speech prosthesis,'' {\em Frontiers in neuroscience}, vol.~12, p.~422, 2018.

\bibitem{Collinger2013ChronicImplants}
J.~L. Collinger, B.~Wodlinger, J.~E. Downey, W.~Wang, E.~C. Tyler-Kabara, D.~J. Weber, A.~J.~C. McMorland, M.~Velliste, M.~L. Boninger, and A.~B. Schwartz, ``High-performance neuroprosthetic control by an individual with tetraplegia,'' {\em The Lancet}, vol.~381, pp.~557--564, 2013.

\bibitem{Sussillo2016RNN}
D.~Sussillo, S.~D. Stavisky, J.~C. Kao, S.~I. Ryu, and K.~V. Shenoy, ``Making brain–machine interfaces robust to future neural variability,'' {\em Nature Communications}, vol.~7, p.~13749, 2016.

\bibitem{glaser2020machine}
J.~I. Glaser, A.~S. Benjamin, R.~H. Chowdhury, M.~G. Perich, L.~E. Miller, and K.~P. Kording, ``Machine learning for neural decoding,'' {\em eneuro}, vol.~7, no.~4, pp.~ENEURO--0506, 2020.

\bibitem{defossez2023decoding}
A.~D{\'e}fossez, C.~Caucheteux, J.~Rapin, O.~Kabeli, and J.-R. King, ``Decoding speech perception from non-invasive brain recordings,'' {\em Nature Machine Intelligence}, vol.~5, no.~10, pp.~1097--1107, 2023.

\bibitem{jayalath2024brain}
D.~Jayalath, G.~Landau, B.~Shillingford, M.~Woolrich, and O.~P. Jones, ``The brain's bitter lesson: Scaling speech decoding with self-supervised learning,'' {\em arXiv preprint arXiv:2406.04328}, 2024.

\bibitem{Taylor2002LearningBCI}
D.~M. Taylor, S.~I.~H. Tillery, and A.~B. Schwartz, ``Direct cortical control of 3d neuroprosthetic devices,'' {\em Science}, vol.~296, pp.~1829--1832, 2002.

\bibitem{Ganguly2009Reorganization}
K.~Ganguly and J.~M. Carmena, ``Emergence of a stable cortical map for neuroprosthetic control,'' {\em PLoS Biology}, vol.~7, p.~e1000153, 2009.

\bibitem{luo2022brain}
S.~Luo, Q.~Rabbani, and N.~E. Crone, ``Brain-computer interface: applications to speech decoding and synthesis to augment communication,'' {\em Neurotherapeutics}, vol.~19, no.~1, pp.~263--273, 2022.

\bibitem{silva2024speech}
A.~B. Silva, K.~T. Littlejohn, J.~R. Liu, D.~A. Moses, and E.~F. Chang, ``The speech neuroprosthesis,'' {\em Nature Reviews Neuroscience}, vol.~25, no.~7, pp.~473--492, 2024.

\bibitem{Goldstein2022Shared}
A.~Goldstein, Z.~Zada, E.~Buchnik, M.~Schain, A.~Price, B.~Aubrey, A.~Flinker, W.~Doyle, D.~Friedman, P.~Dugan, L.~Melloni, R.~Reichart, S.~Devore, A.~Flinker, O.~Devinsky, and U.~Hasson, ``Shared computational principles for language processing in humans and deep language models,'' {\em Nature Neuroscience}, vol.~25, pp.~369--380, Mar 2022.

\bibitem{Goldsteinetal2025}
A.~Goldstein, H.~Wang, L.~Niekerken, M.~Schain, Z.~Zada, B.~Aubrey, T.~Sheffer, S.~A. Nastase, H.~Gazula, A.~Singh, {\em et~al.}, ``A unified acoustic-to-speech-to-language embedding space captures the neural basis of natural language processing in everyday conversations,'' {\em Nature human behaviour}, vol.~9, no.~5, pp.~1041--1055, 2025.

\bibitem{evanson2025emergencelanguagedevelopingbrain}
L.~Evanson, C.~Bulteau, M.~Chipaux, G.~Dorfm{\"u}ller, S.~Ferrand-Sorbets, E.~Raffo, S.~Rosenberg, P.~Bourdillon, and J.-R. King, ``Emergence of language in the developing brain,'' {\em arXiv preprint arXiv:2512.05718}, 2025.

\bibitem{gadonneix2026temporalstructurelanguagehierarchy}
J.~Gadonneix, M.~Zhang, J.~Rapin, L.~Evanson, P.~Bourdillon, and J.-R. King, ``Temporal structure of the language hierarchy within small cortical patches,'' {\em arXiv preprint arXiv:2604.03021}, 2026.

\bibitem{zhang2025thought}
M.~Zhang, J.~L{\'e}vy, S.~d'Ascoli, J.~Rapin, F.~Alario, P.~Bourdillon, S.~Pinet, J.-R. King, {\em et~al.}, ``From thought to action: How a hierarchy of neural dynamics supports language production,'' {\em arXiv preprint arXiv:2502.07429}, 2025.

\bibitem{Bouchard2013MotorCortexSpeech}
K.~E. Bouchard, N.~Mesgarani, K.~Johnson, and E.~F. Chang, ``Functional organization of human sensorimotor cortex for speech articulation,'' {\em Nature}, vol.~495, no.~7441, pp.~327--332, 2013.

\bibitem{Cheung2016ArticulatoryCortex}
C.~Cheung, L.~S. Hamilton, K.~Johnson, and E.~F. Chang, ``The auditory representation of speech sounds in human motor cortex,'' {\em eLife}, vol.~5, p.~e12577, 2016.

\bibitem{Mesgarani2014PhoneticFeatures}
N.~Mesgarani, C.~Cheung, K.~Johnson, and E.~F. Chang, ``Phonetic feature encoding in human superior temporal gyrus,'' {\em Science}, vol.~343, no.~6174, pp.~1006--1010, 2014.

\bibitem{Chang2010CategoricalSpeech}
E.~F. Chang, J.~W. Rieger, K.~Johnson, M.~S. Berger, N.~M. Barbaro, and R.~T. Knight, ``Categorical speech representation in human superior temporal gyrus,'' {\em Nature Neuroscience}, vol.~13, no.~11, pp.~1428--1432, 2010.

\bibitem{Friederici2011LanguageNetwork}
A.~D. Friederici, ``The brain basis of language processing: from structure to function,'' {\em Physiological Reviews}, vol.~91, no.~4, pp.~1357--1392, 2011.

\bibitem{Blank2016SpeechHierarchy}
I.~Blank, Z.~Balewski, K.~Mahowald, and E.~Fedorenko, ``Syntactic processing is distributed across the language system,'' {\em Neuroimage}, vol.~127, pp.~307--323, 2016.

\bibitem{metzger2022generalizable}
S.~L. Metzger, J.~R. Liu, D.~A. Moses, M.~E. Dougherty, M.~P. Seaton, K.~T. Littlejohn, J.~Chartier, G.~K. Anumanchipalli, A.~Tu-Chan, K.~Ganguly, {\em et~al.}, ``Generalizable spelling using a speech neuroprosthesis in an individual with severe limb and vocal paralysis,'' {\em Nature communications}, vol.~13, no.~1, p.~6510, 2022.

\bibitem{Hickok2011DualStream}
G.~Hickok, J.~Houde, and F.~Rong, ``Sensorimotor integration in speech processing: computational basis and neural organization,'' {\em Neuron}, vol.~69, no.~3, pp.~407--422, 2011.

\bibitem{Guenther2016NeuralControlSpeech}
F.~H. Guenther, {\em Neural Control of Speech}.
\newblock Cambridge, MA: MIT Press, 2016.

\bibitem{Tourville2008AuditoryFeedback}
J.~A. Tourville, K.~J. Reilly, and F.~H. Guenther, ``Neural mechanisms underlying auditory feedback control of speech,'' {\em Neuroimage}, vol.~39, no.~3, pp.~1429--1443, 2008.

\bibitem{Houde1998SpeechPerturbation}
J.~F. Houde and M.~I. Jordan, ``Sensorimotor adaptation in speech production,'' {\em Science}, vol.~279, no.~5354, pp.~1213--1216, 1998.

\bibitem{Leonard2016DistributedSpeech}
M.~K. Leonard, M.~O. Baud, M.~J. Sjerps, and E.~F. Chang, ``Perceptual restoration of masked speech in human cortex,'' {\em Nature Communications}, vol.~7, p.~13619, 2016.

\bibitem{Huth2016SemanticMaps}
A.~G. Huth, W.~A. de~Heer, T.~L. Griffiths, F.~E. Theunissen, and J.~L. Gallant, ``Natural speech reveals the semantic maps that tile human cerebral cortex,'' {\em Nature}, vol.~532, no.~7600, pp.~453--458, 2016.

\bibitem{Cogan2014TimeResolvedSpeech}
G.~B. Cogan, T.~Thesen, C.~Carlson, W.~Doyle, O.~Devinsky, and B.~Pesaran, ``Sensory-motor transformations for speech occur bilaterally,'' {\em Nature}, vol.~507, no.~7490, pp.~94--98, 2014.

\bibitem{Euston2012TimeScales}
D.~R. Euston, A.~J. Gruber, and B.~L. McNaughton, ``The role of medial prefrontal cortex in memory and decision making,'' {\em Neuron}, vol.~76, no.~6, pp.~1057--1070, 2012.

\bibitem{Friston2010PredictiveCodingSpeech}
K.~Friston, ``The free-energy principle: a unified brain theory?,'' {\em Nature Reviews Neuroscience}, vol.~11, no.~2, pp.~127--138, 2010.

\bibitem{Gal2016DropoutUncertainty}
Y.~Gal and Z.~Ghahramani, ``Dropout as a bayesian approximation: Representing model uncertainty in deep learning,'' in {\em International Conference on Machine Learning (ICML)}, pp.~1050--1059, 2016.

\bibitem{Hochberg2012IntracorticalBCI}
L.~R. Hochberg, D.~Bacher, B.~Jarosiewicz, N.~Y. Masse, J.~D. Simeral, J.~Vogel, S.~Haddadin, J.~Liu, S.~S. Cash, P.~van~der Smagt, and J.~P. Donoghue, ``Reach and grasp by people with tetraplegia using a neurally controlled robotic arm,'' {\em Nature}, vol.~485, no.~7398, pp.~372--375, 2012.

\bibitem{davidoff2020agency}
E.~J. Davidoff, ``Agency and accountability: ethical considerations for brain-computer interfaces,'' {\em The Rutgers journal of bioethics}, vol.~11, p.~9, 2020.

\bibitem{musk2019integrated}
E.~Musk {\em et~al.}, ``An integrated brain-machine interface platform with thousands of channels,'' {\em Journal of medical Internet research}, vol.~21, no.~10, p.~e16194, 2019.

\bibitem{Neuralink2024PRIME}
{Neuralink}, ``Prime study progress update,'' Apr. 2024.
\newblock Published April 12, 2024.

\bibitem{mitchell2023assessment}
P.~Mitchell, S.~C. Lee, P.~E. Yoo, A.~Morokoff, R.~P. Sharma, D.~L. Williams, C.~MacIsaac, M.~E. Howard, L.~Irving, I.~Vrljic, {\em et~al.}, ``Assessment of safety of a fully implanted endovascular brain-computer interface for severe paralysis in 4 patients: the stentrode with thought-controlled digital switch (switch) study,'' {\em JAMA neurology}, vol.~80, no.~3, pp.~270--278, 2023.

\bibitem{hettick2025minimally}
M.~Hettick, E.~Ho, A.~J. Poole, M.~Monge, D.~Papageorgiou, {\em et~al.}, ``Minimally invasive implantation of scalable high-density cortical microelectrode arrays for multimodal neural decoding and stimulation,'' {\em Nature Biomedical Engineering}, 2025.

\bibitem{schroter2025advances}
M.~Schr{\"o}ter, F.~Cardes, C.-V.~H. Bui, L.~D. Dodi, T.~G{\"a}nswein, J.~Bartram, L.~Sadiraj, P.~Hornauer, S.~Kumar, M.~Pascual-Garcia, {\em et~al.}, ``Advances in large-scale electrophysiology with high-density microelectrode arrays,'' {\em Lab on a Chip}, vol.~25, no.~19, pp.~4844--4885, 2025.

\bibitem{Leuthardt2004ECoG}
E.~C. Leuthardt, G.~Schalk, J.~R. Wolpaw, J.~G. Ojemann, and D.~W. Moran, ``A brain--computer interface using electrocorticographic signals in humans,'' {\em Journal of Neural Engineering}, vol.~1, no.~2, pp.~63--71, 2004.

\bibitem{fifer2013simultaneous}
M.~S. Fifer, G.~Hotson, B.~A. Wester, D.~P. McMullen, Y.~Wang, M.~S. Johannes, K.~D. Katyal, J.~B. Helder, M.~P. Para, R.~J. Vogelstein, {\em et~al.}, ``Simultaneous neural control of simple reaching and grasping with the modular prosthetic limb using intracranial eeg,'' {\em IEEE transactions on neural systems and rehabilitation engineering}, vol.~22, no.~3, pp.~695--705, 2013.

\bibitem{Crone2001HighGamma}
N.~E. Crone, D.~L. Miglioretti, B.~Gordon, and R.~P. Lesser, ``Functional mapping of human sensorimotor cortex with electrocorticographic spectral analysis. ii. event-related synchronization in the gamma band,'' {\em Brain}, vol.~121, no.~12, pp.~2301--2315, 1998.

\bibitem{Buzsaki2012SEEG}
G.~Buzs{\'a}ki, C.~A. Anastassiou, and C.~Koch, ``The origin of extracellular fields and currents — eeg, ecog, lfp and spikes,'' {\em Nature Reviews Neuroscience}, vol.~13, no.~6, pp.~407--420, 2012.

\bibitem{li2018optimal}
G.~Li, S.~Jiang, S.~E. Paraskevopoulou, M.~Wang, Y.~Xu, Z.~Wu, L.~Chen, D.~Zhang, and G.~Schalk, ``Optimal referencing for stereo-electroencephalographic (seeg) recordings,'' {\em NeuroImage}, vol.~183, pp.~327--335, 2018.

\bibitem{khoo2020technical}
H.~M. Khoo, J.~A. Hall, F.~Dubeau, N.~Tani, S.~Oshino, Y.~Fujita, J.~Gotman, and H.~Kishima, ``Technical aspects of seeg and its interpretation in the delineation of the epileptogenic zone,'' {\em Neurologia medico-chirurgica}, vol.~60, no.~12, pp.~565--580, 2020.

\bibitem{Flinker2015SpeechSEEG}
A.~Flinker, A.~Korzeniewska, A.~Y. Shestyuk, P.~J. Franaszczuk, N.~F. Dronkers, R.~T. Knight, and N.~E. Crone, ``Redefining the role of broca's area in speech,'' {\em Proceedings of the National Academy of Sciences}, vol.~112, no.~9, pp.~2871--2875, 2015.

\bibitem{Chartier2018SEEGSpeech}
J.~Chartier, G.~K. Anumanchipalli, K.~Johnson, and E.~F. Chang, ``Encoding of articulatory kinematic trajectories in human speech sensorimotor cortex,'' {\em Neuron}, vol.~98, no.~5, pp.~1042--1054, 2018.

\bibitem{Willett2021Intracortical}
F.~R. Willett, D.~T. Avansino, L.~R. Hochberg, J.~M. Henderson, and K.~V. Shenoy, ``High-performance brain-to-text communication via handwriting,'' {\em Nature}, vol.~593, no.~7858, pp.~249--254, 2021.

\bibitem{Barrese2016FailureModes}
J.~C. Barrese, N.~Rao, K.~Paroo, C.~Triebwasser, C.~Vargas-Irwin, L.~Franquemont, and J.~P. Donoghue, ``Failure mode analysis of silicon-based intracortical microelectrode arrays in non-human primates,'' {\em Journal of Neural Engineering}, vol.~10, no.~6, p.~066014, 2013.

\bibitem{Lalor2009EEGSpeech}
E.~C. Lalor, A.~J. Power, R.~B. Reilly, and J.~J. Foxe, ``Resolving precise temporal processing properties of the auditory system using continuous stimuli,'' {\em Journal of Neurophysiology}, vol.~102, no.~1, pp.~349--359, 2009.

\bibitem{cooney2021bimodal}
C.~Cooney, R.~Folli, and D.~Coyle, ``A bimodal deep learning architecture for eeg-fnirs decoding of overt and imagined speech,'' {\em IEEE Transactions on Biomedical Engineering}, vol.~69, no.~6, pp.~1983--1994, 2021.

\bibitem{gwilliams2023introducing}
L.~Gwilliams, G.~Flick, A.~Marantz, L.~Pylkk{\"a}nen, D.~Poeppel, and J.-R. King, ``Introducing meg-masc a high-quality magneto-encephalography dataset for evaluating natural speech processing,'' {\em Scientific data}, vol.~10, no.~1, p.~862, 2023.

\bibitem{d2025towards}
S.~d’Ascoli, C.~Bel, J.~Rapin, H.~Banville, Y.~Benchetrit, C.~Pallier, and J.-R. King, ``Towards decoding individual words from non-invasive brain recordings,'' {\em Nature Communications}, vol.~16, no.~1, p.~10521, 2025.

\bibitem{ozdogan2026libribrain}
M.~{\"O}zdogan, G.~Landau, G.~Elvers, D.~Jayalath, P.~Somaiya, F.~Mantegna, M.~Woolrich, and O.~Parker~Jones, ``Libribrain: Over 50 hours of within-subject meg to improve speech decoding methods at scale,'' {\em Advances in Neural Information Processing Systems}, vol.~38, 2026.

\bibitem{hollenstein2018zuco}
N.~Hollenstein, J.~Rotsztejn, M.~Troendle, A.~Pedroni, C.~Zhang, and N.~Langer, ``Zuco, a simultaneous eeg and eye-tracking resource for natural sentence reading,'' {\em Scientific data}, vol.~5, no.~1, p.~180291, 2018.

\bibitem{jo2025evaluating}
H.~Jo, Y.~Yang, J.~Han, Y.~Duan, H.~Xiong, and W.~H. Lee, ``Evaluating eeg-to-text models through noise-based performance analysis,'' {\em Scientific Reports}, 2025.

\bibitem{antonello2024evidence}
R.~Antonello and E.~Cheng, ``Evidence from fmri supports a two-phase abstraction process in language models,'' in {\em UniReps: 2nd Edition of the Workshop on Unifying Representations in Neural Models}, 2024.

\bibitem{herff2020potential}
C.~Herff, D.~J. Krusienski, and P.~Kubben, ``The potential of stereotactic-eeg for brain-computer interfaces: current progress and future directions,'' {\em Frontiers in neuroscience}, vol.~14, p.~123, 2020.

\bibitem{Kupers2024}
E.~R. Kupers, T.~Knapen, E.~P. Merriam, and K.~N. Kay, ``Principles of intensive human neuroimaging,'' {\em Trends in Neurosciences}, vol.~47, pp.~856--864, Nov 2024.
\newblock Epub 2024 Oct 24.

\bibitem{Degenhart2020Stability}
A.~D. Degenhart, W.~E. Bishop, E.~R. Oby, E.~C. Tyler-Kabara, S.~M. Chase, A.~P. Batista, and B.~M. Yu, ``Stabilization of a brain–computer interface via the alignment of low-dimensional spaces of neural activity,'' {\em Nature Biomedical Engineering}, vol.~4, no.~7, pp.~672--685, 2020.

\bibitem{felton2012mental}
E.~A. Felton, J.~C. Williams, G.~C. Vanderheiden, and R.~G. Radwin, ``Mental workload during brain--computer interface training,'' {\em Ergonomics}, vol.~55, no.~5, pp.~526--537, 2012.

\bibitem{Wolpaw2018HomeUseNeurology}
J.~R. Wolpaw, R.~S. Bedlack, D.~J. Reda, R.~J. Ringer, P.~G. Banks, T.~M. Vaughan, S.~M. Heckman, L.~M. McCane, C.~S. Carmack, S.~Winden, {\em et~al.}, ``Independent home use of a brain-computer interface by people with amyotrophic lateral sclerosis,'' {\em Neurology}, vol.~91, no.~3, pp.~e258--e267, 2018.

\bibitem{Chartier2018ArticulatoryBCI}
J.~Chartier, G.~K. Anumanchipalli, K.~Johnson, and E.~F. Chang, ``Encoding of articulatory kinematic trajectories in human speech sensorimotor cortex,'' {\em Neuron}, vol.~98, no.~5, pp.~1042--1054, 2018.

\bibitem{Anumanchipalli2019SpeechSynthesis}
G.~K. Anumanchipalli, J.~Chartier, and E.~F. Chang, ``Speech synthesis from neural decoding of spoken sentences,'' {\em Nature}, vol.~568, no.~7753, pp.~493--498, 2019.

\bibitem{Mugler2014PhonemeBCI}
E.~M. Mugler, J.~L. Patton, R.~D. Flint, Z.~A. Wright, S.~U. Schuele, J.~Rosenow, J.~J. Shih, D.~J. Krusienski, and M.~W. Slutzky, ``Direct classification of all american english phonemes using signals from functional speech motor cortex,'' {\em Journal of Neural Engineering}, vol.~11, no.~3, p.~035015, 2014.

\bibitem{tang2017intonational}
C.~Tang, L.~S. Hamilton, and E.~F. Chang, ``Intonational speech prosody encoding in the human auditory cortex,'' {\em Science}, vol.~357, no.~6353, pp.~797--801, 2017.

\bibitem{Wu2006KalmanBCI}
W.~Wu, M.~Black, Y.~Gao, M.~Serruya, A.~Shaikhouni, J.~Donoghue, and E.~Bienenstock, ``Neural decoding of cursor motion using a kalman filter,'' {\em Advances in neural information processing systems}, vol.~15, 2002.

\bibitem{Shenoy2013StateSpaceBCI}
K.~V. Shenoy, M.~Sahani, and M.~M. Churchland, ``Cortical control of arm movements: a dynamical systems perspective,'' {\em Annual Review of Neuroscience}, vol.~36, pp.~337--359, 2013.

\bibitem{Rabiner1989HMM}
L.~R. Rabiner, ``A tutorial on hidden markov models and selected applications in speech recognition,'' {\em Proceedings of the IEEE}, vol.~77, no.~2, pp.~257--286, 1989.

\bibitem{Schirrmeister2017DeepEEG}
R.~T. Schirrmeister, J.~T. Springenberg, L.~D.~J. Fiederer, M.~Glasstetter, K.~Eggensperger, M.~Tangermann, F.~Hutter, W.~Burgard, and T.~Ball, ``Deep learning with convolutional neural networks for eeg decoding and visualization,'' {\em Human Brain Mapping}, vol.~38, no.~11, pp.~5391--5420, 2017.

\bibitem{Lawhern2018EEGNet}
V.~J. Lawhern, A.~J. Solon, N.~R. Waytowich, S.~M. Gordon, C.~P. Hung, and B.~J. Lance, ``{EEGNet}: a compact convolutional neural network for eeg-based brain--computer interfaces,'' {\em Journal of Neural Engineering}, vol.~15, no.~5, p.~056013, 2018.

\bibitem{Hochreiter1997LSTM}
S.~Hochreiter and J.~Schmidhuber, ``Long short-term memory,'' {\em Neural Computation}, vol.~9, no.~8, pp.~1735--1780, 1997.

\bibitem{Bai2018TCN}
S.~Bai, J.~Z. Kolter, and V.~Koltun, ``An empirical evaluation of generic convolutional and recurrent networks for sequence modeling,'' in {\em International Conference on Learning Representations (ICLR)}, 2018.

\bibitem{graves2006connectionist}
A.~Graves, S.~Fern{\'a}ndez, F.~Gomez, and J.~Schmidhuber, ``Connectionist temporal classification: labelling unsegmented sequence data with recurrent neural networks,'' in {\em Proceedings of the 23rd international conference on Machine learning}, pp.~369--376, 2006.

\bibitem{Vaswani2017Attention}
A.~Vaswani, N.~Shazeer, N.~Parmar, J.~Uszkoreit, L.~Jones, A.~N. Gomez, {\L}.~Kaiser, and I.~Polosukhin, ``Attention is all you need,'' in {\em Advances in Neural Information Processing Systems}, vol.~30, 2017.

\bibitem{zhangcross}
Y.~Zhang, L.~He, C.~Fan, T.~Liu, H.~Yu, T.~Le, J.~Li, S.~Linderman, L.~Duncker, F.~R. Willett, {\em et~al.}, ``A cross-species neural foundation model for end-to-end speech decoding,'' in {\em The Fourteenth International Conference on Learning Representations}, 2026.

\bibitem{lecun2022path}
Y.~LeCun, ``A path towards autonomous machine intelligence,'' {\em OpenReview}, 2022.

\bibitem{assran2023self}
M.~Assran, Q.~Duval, I.~Misra, P.~Bojanowski, P.~Vincent, M.~Rabbat, Y.~LeCun, and N.~Ballas, ``Self-supervised learning from images with a joint-embedding predictive architecture,'' in {\em Proceedings of the IEEE/CVF Conference on Computer Vision and Pattern Recognition}, pp.~15619--15629, 2023.

\bibitem{chen2024neural}
X.~Chen, R.~Wang, A.~Khalilian-Gourtani, L.~Yu, P.~Dugan, D.~Friedman, W.~Doyle, O.~Devinsky, Y.~Wang, and A.~Flinker, ``A neural speech decoding framework leveraging deep learning and speech synthesis,'' {\em Nature Machine Intelligence}, vol.~6, no.~4, pp.~467--480, 2024.

\bibitem{metzger2023high}
S.~L. Metzger, K.~T. Littlejohn, A.~B. Silva, D.~A. Moses, M.~P. Seaton, R.~Wang, M.~E. Dougherty, J.~R. Liu, P.~Wu, M.~A. Berger, {\em et~al.}, ``A high-performance neuroprosthesis for speech decoding and avatar control,'' {\em Nature}, vol.~620, no.~7976, pp.~1037--1046, 2023.

\bibitem{kellis2010decoding}
S.~Kellis, K.~Miller, K.~Thomson, R.~Brown, P.~House, and B.~Greger, ``Decoding spoken words using local field potentials recorded from the cortical surface,'' {\em Journal of neural engineering}, vol.~7, no.~5, p.~056007, 2010.

\bibitem{conant2018human}
D.~F. Conant, K.~E. Bouchard, M.~K. Leonard, and E.~F. Chang, ``Human sensorimotor cortex control of directly measured vocal tract movements during vowel production,'' {\em Journal of Neuroscience}, vol.~38, no.~12, pp.~2955--2966, 2018.

\bibitem{jamali2024semantic}
M.~Jamali, B.~Grannan, J.~Cai, A.~R. Khanna, W.~Mu{\~n}oz, I.~Caprara, A.~C. Paulk, S.~S. Cash, E.~Fedorenko, and Z.~M. Williams, ``Semantic encoding during language comprehension at single-cell resolution,'' {\em Nature}, vol.~631, no.~8021, pp.~610--616, 2024.

\bibitem{stavisky2019neural}
S.~D. Stavisky, F.~R. Willett, G.~H. Wilson, B.~A. Murphy, P.~Rezaii, D.~T. Avansino, W.~D. Memberg, J.~P. Miller, R.~F. Kirsch, L.~R. Hochberg, {\em et~al.}, ``Neural ensemble dynamics in dorsal motor cortex during speech in people with paralysis,'' {\em Elife}, vol.~8, p.~e46015, 2019.

\bibitem{wilson2020decoding}
G.~H. Wilson, S.~D. Stavisky, F.~R. Willett, D.~T. Avansino, J.~N. Kelemen, L.~R. Hochberg, J.~M. Henderson, S.~Druckmann, and K.~V. Shenoy, ``Decoding spoken english from intracortical electrode arrays in dorsal precentral gyrus,'' {\em Journal of neural engineering}, vol.~17, no.~6, p.~066007, 2020.

\bibitem{hochberg2006neuronal}
L.~R. Hochberg, M.~D. Serruya, G.~M. Friehs, J.~A. Mukand, M.~Saleh, A.~H. Caplan, A.~Branner, D.~Chen, R.~D. Penn, and J.~P. Donoghue, ``Neuronal ensemble control of prosthetic devices by a human with tetraplegia,'' {\em Nature}, vol.~442, no.~7099, pp.~164--171, 2006.

\bibitem{simeral2011neural}
J.~Simeral, S.-P. Kim, M.~Black, J.~Donoghue, and L.~Hochberg, ``Neural control of cursor trajectory and click by a human with tetraplegia 1000 days after implant of an intracortical microelectrode array,'' {\em Journal of neural engineering}, vol.~8, no.~2, p.~025027, 2011.

\bibitem{Littlejohn2025StreamingBrainToVoice}
K.~T. Littlejohn, C.~J. Cho, J.~R. Liu, A.~B. Silva, B.~Yu, V.~R. Anderson, C.~M. Kurtz-Miott, S.~Brosler, A.~P. Kashyap, I.~P. Hallinan, A.~Shah, A.~Tu-Chan, K.~Ganguly, D.~A. Moses, E.~F. Chang, and G.~K. Anumanchipalli, ``A streaming brain-to-voice neuroprosthesis to restore naturalistic communication,'' {\em Nature Neuroscience}, 2025.

\bibitem{wairagkar2025instantaneous}
M.~Wairagkar, N.~S. Card, T.~Singer-Clark, X.~Hou, C.~Iacobacci, L.~M. Miller, L.~R. Hochberg, D.~M. Brandman, and S.~D. Stavisky, ``An instantaneous voice-synthesis neuroprosthesis,'' {\em Nature}, vol.~644, no.~8075, pp.~145--152, 2025.

\bibitem{proix2022imagined}
T.~Proix, J.~Delgado~Saa, A.~Christen, S.~Martin, B.~N. Pasley, R.~T. Knight, X.~Tian, D.~Poeppel, W.~K. Doyle, O.~Devinsky, {\em et~al.}, ``Imagined speech can be decoded from low-and cross-frequency intracranial eeg features,'' {\em Nature communications}, vol.~13, no.~1, p.~48, 2022.

\bibitem{dash2020decoding}
D.~Dash, P.~Ferrari, and J.~Wang, ``Decoding imagined and spoken phrases from non-invasive neural (meg) signals,'' {\em Frontiers in neuroscience}, vol.~14, p.~290, 2020.

\bibitem{denby2010silent}
B.~Denby, T.~Schultz, K.~Honda, T.~Hueber, J.~M. Gilbert, and J.~S. Brumberg, ``Silent speech interfaces,'' {\em Speech Communication}, vol.~52, no.~4, pp.~270--287, 2010.

\bibitem{vojtech2021surface}
J.~M. Vojtech, M.~D. Chan, B.~Shiwani, S.~H. Roy, J.~T. Heaton, G.~S. Meltzner, P.~Contessa, G.~De~Luca, R.~Patel, and J.~C. Kline, ``Surface electromyography--based recognition, synthesis, and perception of prosodic subvocal speech,'' {\em Journal of Speech, Language, and Hearing Research}, vol.~64, no.~6S, pp.~2134--2153, 2021.

\bibitem{kubler2009brain}
A.~K{\"u}bler, A.~Furdea, S.~Halder, E.~M. Hammer, F.~Nijboer, and B.~Kotchoubey, ``A brain--computer interface controlled auditory event-related potential (p300) spelling system for locked-in patients,'' {\em Annals of the New York Academy of Sciences}, vol.~1157, no.~1, pp.~90--100, 2009.

\bibitem{pandarinath2017high}
C.~Pandarinath, P.~Nuyujukian, C.~H. Blabe, B.~L. Sorice, J.~Saab, F.~R. Willett, L.~R. Hochberg, K.~V. Shenoy, and J.~M. Henderson, ``High performance communication by people with paralysis using an intracortical brain-computer interface,'' {\em elife}, vol.~6, p.~e18554, 2017.

\bibitem{feghhi2025time}
E.~Feghhi, S.~Kaasyap, N.~Hadidi, and J.~Kao, ``Time-masked transformers with lightweight test-time adaptation for neural speech decoding,'' {\em Advances in Neural Information Processing Systems}, vol.~38, pp.~93047--93072, 2026.

\bibitem{vansteensel2016fully}
M.~J. Vansteensel, E.~G. Pels, M.~G. Bleichner, M.~P. Branco, T.~Denison, Z.~V. Freudenburg, P.~Gosselaar, S.~Leinders, T.~H. Ottens, M.~A. Van Den~Boom, {\em et~al.}, ``Fully implanted brain--computer interface in a locked-in patient with als,'' {\em New England Journal of Medicine}, vol.~375, no.~21, pp.~2060--2066, 2016.

\bibitem{vansteensel2024longevity}
M.~J. Vansteensel, S.~Leinders, M.~P. Branco, N.~E. Crone, T.~Denison, Z.~V. Freudenburg, S.~H. Geukes, P.~H. Gosselaar, M.~Raemaekers, A.~Schippers, {\em et~al.}, ``Longevity of a brain--computer interface for amyotrophic lateral sclerosis,'' {\em New England Journal of Medicine}, vol.~391, no.~7, pp.~619--626, 2024.

\bibitem{oxley2021motor}
T.~J. Oxley, P.~E. Yoo, G.~S. Rind, S.~M. Ronayne, C.~S. Lee, C.~Bird, V.~Hampshire, R.~P. Sharma, A.~Morokoff, D.~L. Williams, {\em et~al.}, ``Motor neuroprosthesis implanted with neurointerventional surgery improves capacity for activities of daily living tasks in severe paralysis: first in-human experience,'' {\em Journal of neurointerventional surgery}, vol.~13, no.~2, pp.~102--108, 2021.

\bibitem{fry2022evaluating}
A.~Fry, H.~W. Chan, N.~Y. Harel, L.~A. Spielman, M.~X. Escalon, and D.~F. Putrino, ``Evaluating the clinical benefit of brain-computer interfaces for control of a personal computer,'' {\em Journal of Neural Engineering}, vol.~19, no.~2, p.~021001, 2022.

\bibitem{sawyer2024digital}
A.~Sawyer, L.~Cooke, N.~F. Ramsey, and D.~Putrino, ``The digital motor output: a conceptual framework for a meaningful clinical performance metric for a motor neuroprosthesis,'' {\em Journal of NeuroInterventional Surgery}, vol.~16, no.~5, pp.~443--446, 2024.

\bibitem{brannigan2024brain}
J.~F. Brannigan, K.~Liyanage, H.~L. Horsfall, L.~Bashford, W.~Muirhead, and A.~Fry, ``Brain--computer interfaces patient preferences: a systematic review,'' {\em Journal of neural engineering}, vol.~21, no.~6, p.~061005, 2024.

\bibitem{dohle2025toward}
E.~Dohle, E.~Swanson, L.~Jovanovic, S.~Yusuf, L.~Thompson, H.~L. Horsfall, W.~Muirhead, L.~Bashford, and J.~Brannigan, ``Toward the clinical translation of implantable brain--computer interfaces for motor impairment: research trends and outcome measures,'' {\em Advanced Science}, vol.~12, no.~32, p.~e01912, 2025.

\bibitem{herff2016automatic}
C.~Herff and T.~Schultz, ``Automatic speech recognition from neural signals: a focused review,'' {\em Frontiers in neuroscience}, vol.~10, p.~429, 2016.

\bibitem{Graves2012RNNTransducer}
A.~Graves, ``Sequence transduction with recurrent neural networks,'' in {\em Proceedings of the 29th International Conference on Machine Learning (ICML) Workshop on Representation Learning}, 2012.

\bibitem{chan2016listen}
W.~Chan, N.~Jaitly, Q.~Le, and O.~Vinyals, ``Listen, attend and spell: A neural network for large vocabulary conversational speech recognition,'' in {\em 2016 IEEE international conference on acoustics, speech and signal processing (ICASSP)}, pp.~4960--4964, IEEE, 2016.

\bibitem{olak2026decoding}
M.~Olak, T.~Boccato, and M.~Ferrante, ``Decoding the decoder: Contextual sequence-to-sequence modeling for intracortical speech decoding,'' {\em arXiv preprint arXiv:2603.20246}, 2026.

\bibitem{Fogg2026generalizable}
Z.~M. Fogg, N.~S. Card, M.~Wairagkar, A.~Srinivasan, T.~Singer-Clark, X.~Hou, E.~Okorokova, H.~Peracha, C.~Iacobacci, T.~Brailow, J.~Jude, H.~Levi-Aharoni, T.~Le, D.~Mifsud, P.~I. Deevi, S.~R. Nason-Tomaszewski, A.~L. Pritchard, Y.~Zhang, B.~Richards, P.~Bechefsky, L.~R. Hochberg, Z.~Williams, K.~Shahlaie, N.~AuYong, D.~B. Rubin, C.~Pandarinath, D.~M. Brandman, and S.~D. Stavisky, ``A generalizable speech neuroprosthesis,'' {\em bioRxiv}, 2026.

\bibitem{zhang2024brain}
D.~Zhang, Z.~Wang, Y.~Qian, Z.~Zhao, Y.~Liu, X.~Hao, W.~Li, S.~Lu, H.~Zhu, L.~Chen, {\em et~al.}, ``A brain-to-text framework for decoding natural tonal sentences,'' {\em Cell Reports}, vol.~43, no.~11, 2024.

\bibitem{orsborn2012closed}
A.~L. Orsborn, S.~Dangi, H.~G. Moorman, and J.~M. Carmena, ``Closed-loop decoder adaptation on intermediate time-scales facilitates rapid bmi performance improvements independent of decoder initialization conditions,'' {\em IEEE Transactions on Neural Systems and Rehabilitation Engineering}, vol.~20, no.~4, pp.~468--477, 2012.

\bibitem{Kunz2025InnerSpeechCell}
E.~M. Kunz, B.~A. Krasa, F.~Kamdar, D.~T. Avansino, N.~Hahn, S.~Yoon, A.~Singh, S.~R. Nason-Tomaszewski, N.~S. Card, J.~J. Jude, {\em et~al.}, ``Inner speech in motor cortex and implications for speech neuroprostheses,'' {\em Cell}, vol.~188, no.~17, pp.~4658--4673, 2025.

\bibitem{Sadtler2014NeuralConstraints}
P.~T. Sadtler, K.~M. Quick, M.~D. Golub, S.~M. Chase, S.~I. Ryu, E.~C. Tyler-Kabara, B.~M. Yu, and A.~P. Batista, ``Neural constraints on learning,'' {\em Nature}, vol.~512, no.~7515, pp.~423--426, 2014.

\bibitem{bhadra2025learning}
K.~Bhadra, A.-L. Giraud, and S.~Marchesotti, ``Learning to operate an imagined speech brain-computer interface involves the spatial and frequency tuning of neural activity,'' {\em Communications Biology}, vol.~8, no.~1, p.~271, 2025.

\bibitem{Gilja2012ReFIT}
V.~Gilja, P.~Nuyujukian, C.~A. Chestek, J.~P. Cunningham, B.~M. Yu, J.~M. Fan, M.~M. Churchland, M.~T. Kaufman, J.~C. Kao, S.~I. Ryu, and K.~V. Shenoy, ``A high-performance neural prosthesis enabled by control algorithm design,'' {\em Nature Neuroscience}, vol.~15, no.~12, pp.~1752--1757, 2012.

\bibitem{lee2025brain}
J.~Y. Lee, S.~Lee, A.~Mishra, X.~Yan, B.~McMahan, B.~Gaisford, C.~Kobashigawa, M.~Qu, C.~Xie, and J.~C. Kao, ``Brain--computer interface control with artificial intelligence copilots,'' {\em Nature machine intelligence}, vol.~7, no.~9, pp.~1510--1523, 2025.

\bibitem{Guo2017Calibration}
C.~Guo, G.~Pleiss, Y.~Sun, and K.~Q. Weinberger, ``On calibration of modern neural networks,'' in {\em International Conference on Machine Learning (ICML)}, pp.~1321--1330, 2017.

\bibitem{card2026long}
N.~S. Card, T.~Singer-Clark, H.~Peracha, C.~Iacobacci, X.~Hou, M.~Wairagkar, Z.~Fogg, E.~C. Offenberg, L.~R. Hochberg, S.~D. Stavisky, {\em et~al.}, ``Long-term independent use of an intracortical brain--computer interface for speech and cursor control,'' {\em Nature medicine}, pp.~1--7, 2026.

\bibitem{madduri2024modeling}
M.~M. Madduri, {\em Modeling and Shaping Human-Machine Interactions in Closed-loop, Co-adaptive Neural Interfaces}.
\newblock University of Washington, 2024.

\bibitem{sankaran2023recommendations}
N.~Sankaran, D.~Moses, W.~Chiong, and E.~F. Chang, ``Recommendations for promoting user agency in the design of speech neuroprostheses,'' {\em Frontiers in Human Neuroscience}, vol.~17, p.~1298129, 2023.

\bibitem{Hart1988NASATLX}
S.~G. Hart and L.~E. Staveland, ``Development of nasa-tlx (task load index): Results of empirical and theoretical research,'' {\em Advances in Psychology}, vol.~52, pp.~139--183, 1988.

\bibitem{Brooke1996SUS}
J.~Brooke, ``Sus: A quick and dirty usability scale,'' {\em Usability Evaluation in Industry}, pp.~189--194, 1996.

\bibitem{perge2013intra}
J.~A. Perge, M.~L. Homer, W.~Q. Malik, S.~Cash, E.~Eskandar, G.~Friehs, J.~P. Donoghue, and L.~R. Hochberg, ``Intra-day signal instabilities affect decoding performance in an intracortical neural interface system,'' {\em Journal of neural engineering}, vol.~10, no.~3, p.~036004, 2013.

\bibitem{sussillo2016making}
D.~Sussillo, S.~D. Stavisky, J.~C. Kao, S.~I. Ryu, and K.~V. Shenoy, ``Making brain--machine interfaces robust to future neural variability,'' {\em Nature communications}, vol.~7, no.~1, p.~13749, 2016.

\bibitem{singh2025transfer}
A.~Singh, T.~Thomas, J.~Li, G.~Hickok, X.~Pitkow, and N.~Tandon, ``Transfer learning via distributed brain recordings enables reliable speech decoding,'' {\em Nature communications}, vol.~16, no.~1, p.~8749, 2025.

\bibitem{Angelopoulos2023Conformal}
A.~N. Angelopoulos and S.~Bates, ``Conformal prediction: A gentle introduction,'' {\em Foundations and Trends in Machine Learning}, vol.~16, no.~4, pp.~494--591, 2023.

\bibitem{FDAImplantedBCIGuidance2021}
{U.S. Food and Drug Administration}, ``Implanted brain--computer interface (bci) devices for patients with paralysis or amputation,'' 2021.

\bibitem{Levinson2016TurnTaking}
S.~C. Levinson, ``Turn-taking in human communication—origins and implications for language processing,'' {\em Trends in Cognitive Sciences}, vol.~20, pp.~6--14, 2016.

\bibitem{LevinsonTorreira2015TimingTurnTaking}
S.~C. Levinson and F.~Torreira, ``Timing in turn-taking and its implications for processing models of language,'' {\em Frontiers in psychology}, vol.~6, p.~731, 2015.

\bibitem{Meyer2023TimingConversation}
A.~S. Meyer, ``Timing in conversation,'' {\em Journal of Cognition}, vol.~6, no.~1, pp.~20--20, 2023.

\bibitem{levin2026cross}
A.~D. Levin, D.~T. Avansino, F.~B. Kamdar, N.~S. Card, M.~Wairagkar, B.~G. Jacques, J.~J. Jude, C.~Iacobacci, B.~E. Lacayo, P.~H. Bechefsky, {\em et~al.}, ``Cross-brain transfer of high-performance intracortical speech and handwriting bcis,'' {\em bioRxiv}, pp.~2026--01, 2026.

\bibitem{Levy2025Brain2Qwerty}
J.~L{\'e}vy, M.~Zhang, S.~Pinet, J.~Rapin, H.~Banville, S.~d'Ascoli, and J.-R. King, ``Brain-to-text decoding: A non-invasive approach via typing,'' {\em arXiv preprint}, 2025.

\bibitem{liu2023decoding}
Y.~Liu, Z.~Zhao, M.~Xu, H.~Yu, Y.~Zhu, J.~Zhang, L.~Bu, X.~Zhang, J.~Lu, Y.~Li, {\em et~al.}, ``Decoding and synthesizing tonal language speech from brain activity,'' {\em Science Advances}, vol.~9, no.~23, p.~eadh0478, 2023.

\bibitem{qian2025real}
Y.~Qian, C.~Liu, P.~Yu, X.~Ran, S.~Li, Q.~Yang, Y.~Liu, L.~Xia, Y.~Wang, J.~Qi, {\em et~al.}, ``Real-time decoding of full-spectrum chinese using brain-computer interface,'' {\em Science Advances}, vol.~11, no.~45, p.~eadz9968, 2025.

\bibitem{silva2024bilingual}
A.~B. Silva, J.~R. Liu, S.~L. Metzger, I.~Bhaya-Grossman, M.~E. Dougherty, M.~P. Seaton, K.~T. Littlejohn, A.~Tu-Chan, K.~Ganguly, D.~A. Moses, {\em et~al.}, ``A bilingual speech neuroprosthesis driven by cortical articulatory representations shared between languages,'' {\em Nature Biomedical Engineering}, vol.~8, no.~8, pp.~977--991, 2024.

\bibitem{guo2025pre}
Y.~Guo, Y.~Dong, M.~K.-P. Ng, and S.~Wang, ``A pre-trained framework for multilingual brain decoding using non-invasive recordings,'' {\em arXiv preprint arXiv:2506.03214}, 2025.

\bibitem{IMDRF_SaMD_N41_2017}
{International Medical Device Regulators Forum}, ``Software as a medical device (samd): Clinical evaluation.'' \url{https://www.imdrf.org}, 2017.
\newblock IMDRF SaMD N41.

\bibitem{FDAEarlyFeasibilityIDE2018}
{U.S. Food and Drug Administration}, ``Investigational device exemptions (ides) for early feasibility medical device clinical studies,'' 2018.

\bibitem{Paradromics2025ConnectOneIDE}
{Paradromics}, ``Paradromics receives fda approval for the connect-one clinical study with the connexus\textsuperscript{\textregistered} brain-computer interface,'' Nov. 2025.

\bibitem{Webb2025MedtechInsightParadromicsIDE}
M.~Webb, ``Paradromics joins neurotech race with fda nod to test bci for restoring speech,'' Dec. 2025.

\bibitem{FDA_Cybersecurity_Guidance_2025}
{U.S. Food and Drug Administration}, ``Cybersecurity in medical devices,'' 2025.

\bibitem{ISO14971_2019}
{International Organization for Standardization}, ``Medical devices --- application of risk management to medical devices,'' 2019.

\bibitem{IMDRF_SaMD_N23_2015}
{International Medical Device Regulators Forum}, ``Software as a medical device (samd): Application of quality management system,'' Tech. Rep. IMDRF/SaMD WG/N23FINAL:2015, International Medical Device Regulators Forum, Oct. 2015.

\bibitem{OECD2019NeurotechRecommendation}
{Organisation for Economic Co-operation and Development}, ``Recommendation of the council on responsible innovation in neurotechnology.'' \url{https://legalinstruments.oecd.org}, 2019.
\newblock OECD/LEGAL/0457.

\bibitem{UNESCO2025NeurotechEthics}
{UNESCO}, ``Recommendation on the ethics of neurotechnology.'' \url{https://www.unesco.org}, 2025.
\newblock Adopted November 2025.

\bibitem{Ruiz2024Neurorights}
S.~Ruiz, L.~Valera, P.~Ramos, and R.~Sitaram, ``Neurorights in the constitution: from neurotechnology to ethics and politics,'' {\em Philosophical Transactions of the Royal Society B: Biological Sciences}, vol.~379, no.~1915, p.~20230098, 2024.

\bibitem{Spichak2025NeurorightsJMIR}
S.~Spichak, ``The controversial push for new brain and neurorights,'' {\em J Med Internet Res}, vol.~27, p.~e72270, 2025.

\bibitem{OECDNeurotechToolkit2024}
{Organisation for Economic Co-operation and Development}, ``Neurotechnology toolkit.'' \url{https://www.oecd.org/sti/neurotechnology.htm}, 2024.
\newblock Policy toolkit and governance resources.

\bibitem{almufareh2025inner}
M.~F. Almufareh, S.~Kausar, M.~Humayun, S.~Tehsin, and A.~Farooq, ``Inner speech decoding: a comprehensive review,'' {\em Wiley Interdisciplinary Reviews: Cognitive Science}, vol.~16, no.~6, p.~e70016, 2025.

\bibitem{Gebru2018Datasheets}
T.~Gebru, J.~Morgenstern, B.~Vecchione, J.~W. Vaughan, H.~Wallach, H.~Daum{\'e}~III, and K.~Crawford, ``Datasheets for datasets,'' {\em Communications of the ACM}, vol.~64, no.~12, pp.~86--92, 2021.

\bibitem{Mitchell2019ModelCards}
M.~Mitchell, S.~Wu, A.~Zaldivar, P.~Barnes, L.~Vasserman, B.~Hutchinson, E.~Spitzer, I.~Raji, and T.~Gebru, ``Model cards for model reporting,'' {\em Proceedings of the Conference on Fairness, Accountability, and Transparency}, pp.~220--229, 2019.

\bibitem{guzman2022chile}
L.~Guzm{\'a}n, ``Chile: Pioneering the protection of neurorights,'' {\em The UNESCO Courier}, vol.~2022, no.~1, pp.~13--14, 2022.

\bibitem{cornejo2023chilean}
I.~Cornejo-Plaza, ``Chilean neurorights legislation and its relevance for mental health: Criticisms and outlook,'' {\em Salud mental}, vol.~46, no.~5, pp.~269--273, 2023.

\end{thebibliography}
\end{document}